\documentclass[aps,pre,showpacs]{revtex4}
\usepackage{amsfonts}
\usepackage{amsmath}
\usepackage{amssymb}
\usepackage{float}
\usepackage{graphicx}
\usepackage{color}

\def\kbar{\protect\@kbar}
\def\@kbar{\relax \bgroup
\def\@tempa{\hbox{\raise.73\ht0
\hbox to0pt{\kern.25\wd0\vrule width.5\wd0 height.1pt
depth.1pt\hss}\box0}}\mathchoice{\setbox0\hbox{$\displaystyle
k$}\@tempa}{\setbox0\hbox{$\textstyle
k$}\@tempa}{\setbox0\hbox{$\scriptstyle
k$}\@tempa}{\setbox0\hbox{$\scriptscriptstyle k$}\@tempa}\egroup}

\begin{document}

\title{\textbf{Floquet Systems with Hall Effect: Topological Properties and Phase Transitions}}
\author{Itzhack Dana and Kazuhiro Kubo}
\affiliation{Department of Physics, Bar-Ilan University, Ramat-Gan 5290002, Israel}

\begin{abstract}
We study the quantum topological properties of Floquet (time-periodic) systems exhibiting Hall effects due to perpendicular magnetic and electric fields. The systems are charged particles periodically kicked by a one-dimensional cosine potential in the presence of such fields, where the magnetic field is perpendicular also to the kicking direction. We consider parameter values including small kicking strength, which enables the quantum evolution to be described by effective Floquet Hamiltonians. We also assume the semiclassical regime of a small scaled Planck constant. In the case of an electric field parallel to the kicking direction, one observes, as the electric-field strength is varied, a series of topological phase transitions, causing changes in the Chern numbers of quasienergy (QE) bands. These transitions are due to band degeneracies which, in an initial wide range of electric-field strengths, are shown to occur when the energy of the classical separatrix is approximately equal to the average QE of the degenerating bands. For larger electric-field strengths, the degeneracies reflect changes in the classical phase-space structure. In the case of an electric field not parallel to the kicking direction and satisfying some resonance conditions, we show that the topological properties of the QE bands are characterized by universal (electric-field independent) Chern numbers. These same numbers also characterize the QE band spectrum in the first case, at the end of the basic electric-field interval where the topological phase transitions take place. The two cases above are known to exhibit significantly different dynamical rates, both classical and quantum. This is well reflected by the different topological properties in the two cases.     
\end{abstract}

\maketitle

\begin{center}
\textbf{I. INTRODUCTION}
\end{center}

The topological characterization of band spectra was initiated in the pioneering paper by Thouless, Kohmoto, Nightingale, and den Nijs (TKNN) \cite{tknn} for the static (time-independent) system of two-dimensional (2D) Bloch electrons perpendicular to a uniform magnetic field. TKNN showed that in the framework of linear-response theory the Hall conductance of this system, for ``rational" magnetic fields and with the Fermi energy in a gap, is quantized as $\sigma e^2/h$, where the integer $\sigma$ is the sum of the Chern numbers associated with the occupied magnetic bands below the gap. TKNN also derived a Diophantine equation determining the values of $\sigma$ for a particular 2D periodic potential under the assumption that the coupling between different Landau levels can be neglected. Later on \cite{daz,dz}, it was shown that the Diophantine equation is just a result of magnetic translational invariance and thus it holds under conditions much more general than those assumed by TKNN, in particular also in the presence of Landau-level coupling \cite{skg}. It was also found that a large class of one-dimensional (1D) quasiperiodic systems exhibit universal Chern numbers uniquely determined by the Diophantine equation \cite{kz,d}. 

If the quantum system is not static but time periodic, one has a Floquet system for which energy spectra are replaced  by quasienergy (QE) ones. The first works showing that QE band spectra can have a nontrivial topological characterization by nonzero Chern numbers considered the so-called ``kicked-Harper" model (KHM) as a simple but typical system \cite{leboeuf,fl,d1,d2,dfw,drf}. The KHM is exactly equivalent to the system of periodically kicked charged particles in a uniform magnetic field for some parameter values \cite{d3}. In a semiclassical regime of the KHM, zero or nonzero values of the Chern number of a QE band are usually associated with QE eigenstates that are localized or extended in phase space, respectively \cite{leboeuf,d2,drf}.    

More recently \cite{fti,lsz,jg1,jg2,jg3,d4}, there has been a considerable new interest in the topological properties of Floquet systems, especially in the context of in-gap edge states and Floquet topological insulators \cite{fti,lsz,jg3}. We mention here, in particular, an exact quantum-transport meaning for the Chern numbers of a class of Floquet systems depending periodically on a phase parameter $\alpha$ \cite{jg1,jg2,jg3,d4}. For these systems, the Chern number of a QE band determines the change in the mean momentum of a state in the band when $\alpha$ is adiabatically varied by one period ($2\pi$) \cite{jg1}. These systems, like the KHM, exhibit translational invariance in a phase space which allows to derive a Diophantine equation for the Chern numbers of both these systems \cite{d4} and the KHM \cite{d2}. This equation is analogous to that for the TKNN static system, with the magnetic translational invariance being essentially a translational invariance in a phase space \cite{dz}. 

In this paper, we study the topological properties of a different class of Floquet systems having phase-space translational invariance. These systems are described by the general quantum Hamiltonian   
\begin{equation}\label{GH}
\hat{H}=\frac{\mathbf{\hat{\Pi }}^{2}}{2m_{\rm e}}+e\mathbf{E\cdot \hat{r}}+\kappa V(\hat{x},t), 
\end{equation}
where $m_{\rm e}$ and $-e$ are the effective mass and charge of an electron, $\mathbf{\hat{\Pi}}= \mathbf{\hat{p}}+e\mathbf{B}\times{\mathbf{\hat{r}}}/(2c)$ is the kinetic momentum for a uniform magnetic field $\mathbf{B}$ in the $z$ direction, $\mathbf{E}$ is a uniform electric field in the $(x,y)$ plane, and $\kappa V(\hat{x},t)$ is a 1D potential of strength $\kappa$, periodic in both $x$ and $t$ with period $2\pi$ and $T$, respectively. We shall assume, in most of this paper, that $V(\hat{x},t)$ is the ``kicking" potential in Eq. (\ref{Vk}) below, with the time dependence given by a periodic delta function. The system (\ref{GH}), with this time dependence and with $\mathbf{E}$ in the $y$ direction, is the ``kicked Hall system" (KHS), introduced in its classical version and studied in Ref. \cite{bhd}. Quantum properties of the KHS were investigated in Ref. \cite{dk}. For the KHS, the perpendicular $\mathbf{B}$ and $\mathbf{E}$ fields lead to a Hall current in the kicking ($x$) direction which, for small $\kappa$, was found to cause a significant suppression of classical and quantum dynamical rates \cite{bhd,dk} relative to these rates in the case of $\mathbf{E}=0$ \cite{dd1} (see also the end of Sec. II C). As shown in Sec. II B, the latter case is essentially equivalent to that of nonzero $\mathbf{E}$ in the $x$ direction. Similarly, the system for $E_x=0,\ E_y\neq 0$, i.e., the KHS, is equivalent to the general system with $E_x,\ E_y\neq 0$.

Our results hold in the semiclassical regime of a small scaled Planck constant. We shall also assume small $\kappa$, so that the quantum evolution and QE spectra can be approximately described by effective Floquet Hamiltonians corresponding to static systems. We then show that the dynamical difference above between the cases of $E_y\neq 0$ and $E_y=0$ is clearly reflected in the topological properties in the two cases. In the case of $E_y=0$ ($\mathbf{E}$ in the $x$ direction), we observe, as $E_x$ is varied, a series of topological phase transitions causing changes in the Chern numbers of QE bands. These transitions are due to band degeneracies which, in an initial wide range of $E_x$, are shown to occur when the energy of the classical separatrix is approximately equal to the average QE of the degenerating bands. For larger $E_x$, the degeneracies reflect significant changes in the classical phase-space structure. In the case of nonzero $E_y$ satisfying some resonance conditions, we show that the topological properties of the QE bands for sufficiently small $\kappa$ are characterized by universal (electric-field independent) Chern numbers. These same numbers also characterize the QE bands in the case above of $E_y=0$, at the end of the basic interval of $E_x$ where the topological phase transitions take place. Our analysis uses, among other things, a Diophantine equation for the system following from the phase-space translational invariance of the basic evolution operator. 

The paper is organized as follows. In Sec. II, we express the system (\ref{GH}) in natural variables and present the basic evolution operator in some time period in cases of $E_y\neq 0$ and $E_y=0$. The leading terms in an expansion of the effective Hamiltonian for small $\kappa$ are written in these cases. In Sec. III, we present relevant facts about QE bands, Chern numbers, and Diophantine equations for systems with translational invariance in a phase space. In Sec. IV, we study the topological properties of the system (\ref{GH}) in a semiclassical regime in the cases of $E_y=0$ and $E_y\neq 0$. As $E_x$ is varied, topological phase transitions are shown to occur for $E_y=0$ but not for nonzero $E_y$ satisfying some resonance conditions. In Sec. V, we examine how the topological phase transitions occurring for $E_y=0$ are manifested in the case that the small scaled Planck constant takes irrational values or rational values different from those assumed in Sec. IV. A summary and conclusions are presented in Sec. VI. Several technical details are brought in the Appendices.

\begin{center}
\textbf{II. SYSTEMS IN NATURAL VARIABLES}
\end{center}

\begin{center}
\textbf{A. Hamiltonians}
\end{center}
 
Without loss of generality, we choose units such that $m_{\rm e}=e=1$ in Eq. (\ref{GH}). Let us express (\ref{GH}) in the two natural degrees of freedom in a magnetic field \cite{jl}, given by the independent pairs of conjugate variables $(\hat{x}_{\mathrm{c}},\hat{y}_{\mathrm{c}})$ (coordinates of the cyclotron-orbit center) and $(\hat{u}=-\hat{\Pi} _{x}/\omega ,\hat{v}=\hat{\Pi} _{y}/\omega )$, where $\omega =B/c$ is the cyclotron angular velocity. We can still choose units such that also $\omega =1$; one then has the commutation relations \cite{jl}
\begin{align}\label{cr}
[\hat{u},\hat{v}] &= [\hat{x}_{\mathrm{c}},\hat{y}_{\mathrm{c}}]=i\hbar , \notag \\
[\hat{u},\hat{x}_{\mathrm{c}}]=[\hat{v},\hat{x}_{\mathrm{c}}] &= [\hat{u},\hat{y}_{\mathrm{c}}]=[\hat{v},\hat{y}_{\mathrm{c}}]=0,
\end{align} 
where $\hbar$ denotes a scaled Planck constant in the above units (such that $m_{\rm e}=e=\omega =1$). From simple geometry, one has $\hat{x}=\hat{x}_{\mathrm{c}}+\hat{v}$ and $\hat{y}=\hat{y}_{\mathrm{c}}+\hat{u}$. The Hamiltonian (\ref{GH}) then assumes the form
\begin{equation}\label{eKHO}
\hat{H}=\frac{1}{2}(\hat{u}^{2}+\hat{v}^{2})+E_x(\hat{x}_{\mathrm{c}}+\hat{v})
+E_y(\hat{y}_{\mathrm{c}}+\hat{u})
+\kappa V(\hat{x}_{\mathrm{c}}+\hat{v},t).  
\end{equation}
We note that the first term on the right-hand side of Eq. (\ref{eKHO}) describes a harmonic oscillator in the conjugate variables $(\hat{u},\hat{v})$. Equation (\ref{eKHO}) can be simplified by defining $\hat{u}'=\hat{u}+E_y$ and $\hat{v}'=\hat{v}+E_x$. After neglecting insignificant constants $-E_x^2/2$ and $-E_y^2/2$ and re-denoting $(\hat{u}',\hat{v}')$ by $(\hat{u},\hat{v})$, we get
\begin{equation}\label{eKHOs}
\hat{H}=\frac{1}{2}(\hat{u}^{2}+\hat{v}^{2})+E_x\hat{x}_{\mathrm{c}}+E_y\hat{y}_{\mathrm{c}}
+\kappa V(\hat{x}_{\mathrm{c}}-E_x+\hat{v},t).  
\end{equation}
From Eqs. (\ref{cr}), (\ref{eKHOs}), and Heisenberg equation, one has $d\hat{x}_{\mathrm{c}}/dt=i[\hat{H},\hat{x}_{\mathrm{c}}]/\hbar =E_y$ and, for small $\kappa\ll 1$, $d\hat{y}_{\mathrm{c}}/dt=i[\hat{H},\hat{y}_{\mathrm{c}}]/\hbar =-E_x+ O(\kappa )$; thus,
\begin{equation}\label{xct}
\hat{x}_{\mathrm{c}}(t)=\hat{x}_{\mathrm{c}}(0)+E_yt .
\end{equation}
\begin{equation}\label{yct}
\hat{y}_{\mathrm{c}}(t)=\hat{y}_{\mathrm{c}}(0)-E_xt+O(\kappa ) .
\end{equation}
Equations (\ref{xct}) and (\ref{yct}) express an approximate Hall effect for small $\kappa$, i.e., a motion of the cyclotron orbit center $(\hat{x}_{\mathrm{c}},\hat{y}_{\mathrm{c}})$ with an almost constant velocity $(E_y,-E_x)$ perpendicular to both $\mathbf{B}$ and $\mathbf{E}$. However, the Hall effect for $\hat{x}_{\mathrm{c}}$, i.e., Eq. (\ref{xct}), is an exact one. In particular, for $E_y=0$, $\hat{x}_{\mathrm{c}}$ is an exact constant of the motion. 

\begin{center}
\textbf{B. Basic evolution operator}
\end{center}

For the sake of simplicity and definiteness, we shall make from now on some assumptions about the system (\ref{eKHOs}), besides the assumption of small $\kappa$. First, we assume that the potential $V(\hat{x},t)$ has the simple periodic dependence $-\cos (\hat{x})$ on $\hat{x}$ and that its periodic dependence on $t$ has a very broad frequency spectrum, so that this dependence can be approximated by a periodic delta function. Thus, $V(\hat{x},t)$ is a ``kicking" potential:
\begin{equation}\label{Vk}
V(\hat{x},t) =-\cos (\hat{x})\sum_{s=-\infty}^{\infty}\delta (t-sT). 
\end{equation}
Second, we assume resonance conditions satisfied by the parameters $\gamma =\omega T=T$ and $\eta =E_yT$:
\begin{equation}\label{rc}
\gamma =\omega T=\frac{\pi}{2},\ \ \ \ \ \eta =E_yT=2\pi \frac{w}{\ell},
\end{equation}
where $(w,\ell )$ are coprime integers.

The evolution of a quantum state $\left|\Psi (t)\right\rangle$ in one time period $T$, say from $t=-0$ (i.e., $t$ infinitesimally close from the left to $t=0$) to $t=T-0$, is determined by the evolution operator $\hat{U}$, $\hat{U}\left|\Psi (-0)\right\rangle = \left|\Psi (T-0)\right\rangle$. An expression for $\hat{U}$ is given in Appendix A, see Eq. (\ref{U1}); this expression generalizes that in special cases considered in previous works \cite{dd1,dk} (see also below),

Under the resonance condition $\gamma =\pi /2$ in Eq. (\ref{rc}), one cyclotron period equals $4T$, so that this resonance is realized in four kicking periods. The resonance condition on $\eta$ in Eq. (\ref{rc}) means that the variable $\hat{x}_{\rm c}(t)$ in Eq. (\ref{xct}) will cover an integer number $w$ of $2\pi$ spatial periods of the potential (\ref{Vk}) after a minimal time $\ell T$. Thus, the two resonances will both be realized after a minimal time $rT$, where $r={\rm lcm}(4,\ell )$, the least common multiple of $(4,\ell )$. The basic evolution operator is then $\hat{U}^r$. As shown in Appendix A, one has, up to nonrelevant terms,
\begin{equation}\label{Uper}
\hat{U}^r=\prod_{j=1}^r \exp\left[i\mu \cos\left(x_{\mathrm{c}}-E_x-j\eta
+\hat{v}_j\right)\right] . 
\end{equation}
Here $\mu =\kappa /\hbar$, $x_{\rm c}$ is an arbitrary constant (an eigenvalue of $\hat{x}_{\rm c}$), and the factors in the product are arranged from left to right in order of increasing $j$ after defining $\hat{v}_1=\hat{u}$, $\hat{v}_2=-\hat{v}$, $\hat{v}_3=-\hat{u}$, $\hat{v}_4=\hat{v}$, with $\hat{v}_j$ being periodic in $j$ with period $4$. In particular, for $E_y=0$ ($\eta =0=2\pi 0/1$) with $r=4$, Eq. (\ref{Uper}) reduces to
\begin{align}\label{Upa4}
\hat{U}^4 &=\exp [i\mu\cos (x_{{\rm c}}-E_x+\hat{u})]\exp [i\mu\cos (x_{{\rm c}}-E_x-\hat{v})] \notag \\
&\times \exp [i\mu\cos (x_{{\rm c}}-E_x-\hat{u})]\exp [i\mu\cos (x_{{\rm c}}-E_x+\hat{v})]. 
\end{align}
Since $x_{{\rm c}}$ appears just as an additive constant to $E_x$ in Eqs. (\ref{Uper}) and (\ref{Upa4}), the case of $E_x=0$ for all $x_{{\rm c}}$ is clearly equivalent to the case of $x_{{\rm c}}=0$ for all $E_x$. Thus, the latter case for $E_y=0$ is essentially the same as that of the system without electric field ($\mathbf{E}=0$) for all $x_{{\rm c}}$, whose quntum-dynamical properties were studied in Ref. \cite{dd1}. We shall be mainly interested in the dependence on $E_x$, so that we shall choose from now on $x_{{\rm c}}=0$ and consider all $E_x$.

\begin{center}
\textbf{C. Effective Hamiltonians and dynamics}
\end{center}
  
Being unitary, the evolution operator (\ref{Uper}) can be formally expresed as $\exp [-i\mu \hat{H}^{(\rm e)}]$, where $\hat{H}^{(\rm e)}$ is a Hermitian operator called the effective Hamiltonian. We present here the relevant results about $\hat{H}^{(\rm e)}$. The main lines of the derivation of these results, similar to those in works \cite{dd1,dk}, are given in Appendix A. In general, $\hat{H}^{(\rm e)}$ can be formally expanded as follows:
\begin{equation}\label{Hee}
\hat{H}^{(\rm e)}(\hat{u},\hat{v})=\sum_{\imath =0}^{\infty}\epsilon ^{\imath}\hat{H}_{\imath }^{(\rm e)}(\hat{u},\hat{v}),
\end{equation}
where
\begin{equation}\label{eps}
\epsilon =\mu \sin (\pi\hbar_{\rm s}) = \frac{\kappa}{2}\frac{\sin (\pi\hbar_{\rm s})}{\pi\hbar_{\rm s}}
\end{equation}
and $\hbar_{\rm s}=\hbar /(2\pi )$. Like the evolution operators (\ref{Uper}) and (\ref{Upa4}), the effective Hamiltonian (\ref{Hee}) and all its coefficients $\hat{H}_{\imath}^{(\rm e)}(\hat{u},\hat{v})$ are translationally invariant in the $(u,v)$ phase space with a $2\pi\times 2\pi$ unit cell. Let us write the first two terms in the expansion (\ref{Hee}) in different cases, choosing $x_{{\rm c}}=0$ as explained at the end of Sec. II B. In the case of the operator (\ref{Upa4}) ($\eta =0$, $r=4$), one has
\begin{eqnarray}
\hat{H}_0^{(\rm e)}(\hat{u},\hat{v})&=&-2\cos (E_x)[\cos (\hat{u}) + \cos (\hat{v})] , \label{HeQMEta0a}\\
\hat{H}_1^{(\rm e)}(\hat{u},\hat{v})
&=& -\cos(\hat{u}+\hat{v})+\cos(2E_x)\cos(\hat{u}-\hat{v}) \notag \\
&~&-\sin(2E_x)\sin(\hat{u}+\hat{v}). \label{HeQMEta0b}
\end{eqnarray}

In the case of the operator (\ref{Uper}), we shall restrict our attention to generic values of $r={\rm lcm}(4,\ell )>8$, i.e., the special values of $r=4,8$ will not be considered. We find, for $r>8$,
\begin{align}
\hat{H}_0^{(\rm e)}(\hat{u},\hat{v})&=0,
\label{HeQMEtaNon0a} \\
\hat{H}_1^{(\rm e)}(\hat{u},\hat{v})
&=-\frac{r}{8\cos (\eta)}\big[ \cos(\hat{u}+\hat{v})+\cos(\hat{u}-\hat{v})\big]  \label{HeQMEtaNon0b}.  
\end{align}
One should note that, unlike Eqs. (\ref{HeQMEta0a}) and (\ref{HeQMEta0b}), Eqs. (\ref{HeQMEtaNon0a}) and (\ref{HeQMEtaNon0b}) are completely independent of $E_x$. We also remark that for $E_x=\pi /2$ Eqs. (\ref{HeQMEta0a}) and (\ref{HeQMEta0b}) coincide, up to a constant factor, with Eqs. (\ref{HeQMEtaNon0a}) and (\ref{HeQMEtaNon0b}). In Appendix B, we give a direct derivation of the classical analogs of Eqs. (\ref{HeQMEta0a}) and (\ref{HeQMEta0b}); see Eq. (\ref{HeEta0}) that includes also the second-order classical term $H_2^{(\rm e)}(u,v)$.

The results above have significant implications on the dynamics of the system for small $\kappa$. In the case of $\eta\neq 0$ and $r={\rm lcm}(4,\ell )>8$, with Eqs. (\ref{HeQMEtaNon0a}) and (\ref{HeQMEtaNon0b}), the leading term in the effective Hamiltonian (\ref{Hee}) is of order $\epsilon$ or $\kappa$ [see Eq. (\ref{eps})]. For $\eta =0$, on the other hand, with Eqs. (\ref{HeQMEta0a}) and (\ref{HeQMEta0b}), the leading term is of order $O(1)$. As a consequence, for $\eta\neq 0$, the quantum evolution of expectation values is much slower than that in the case of $\eta=0$ \cite{dk}. This is a quantum analog of the ``superweak-chaos" phenomenon in the corresponding classical systems \cite{bhd}: For $\eta\neq 0$, the chaotic diffusion is much slower than that for $\eta =0$. 

\begin{center}
\textbf{III. QE BANDS AND CHERN NUMBERS}
\end{center}

We present here relevant known facts \cite{d2,dk,dd1} about the exact QE bands and eigenstates of phase-space translationally invariant evolution operators, such as (\ref{Uper}), and the associated topological Chern numbers. QEs are the phases ${\cal E}$ determining the eigenvalues $\exp(-i{\cal E} )$ of the basic evolution operator (\ref{Uper}): $\hat{U}^r|\Psi_{\cal E}\rangle =\exp(-i{\cal E} ) |\Psi_{\cal E}\rangle$. Alternatively, ${\cal E}/\mu$ can be viewed as an ``energy" eigenvalue of the static effective Hamiltonian (\ref{Hee}). Clearly, $\hat{U}^r$ or $\hat{H}^{(\rm e)}(\hat{u},\hat{v})$ commute with translations by $2\pi$ in $\hat{u}$ and $\hat{v}$. Since $[\hat{u},\hat{v}]=i\hbar=2\pi i\hbar_{\rm s}$, one has $\hat{u}=2\pi i\hbar_{\rm s}d/dv$ and $\hat{v}=-2\pi i\hbar_{\rm s}d/du$, so that the translations above are given by the operators $\hat{D}_0 =\exp(i\hat{v}/\hbar_{\rm s})$ and $\hat{D}_1 =\exp(-i\hat{u}/\hbar_{\rm s})$. In general, the latter operators do not commute. However, for rational $\hbar_{\rm s}=q/p$, where $q$ and $p$ are coprime integers, the operators
\begin{equation}\label{D12} 
\hat{D}_1=\exp(-i\hat{u}/\hbar_{\rm s}),\ \ \ \hat{D}_2=\hat{D}_0^q=\exp(ip\hat{v}),
\end{equation} 
commute and, of course, they commute also with $\hat{U}^r$ or $\hat{H}^{(\rm e)}(\hat{u},\hat{v})$. The simultaneous eigenstates of $\hat{D}_1$, $\hat{D}_2$, and $\hat{U}^r$ or $\hat{H}^{(\rm e)}(\hat{u},\hat{v})$ in the $v$-representation can be written as \cite{d2,dk,dd1,note}:
\begin{eqnarray}\label{qes}
\langle v|\Psi_{b,\mathbf{k}}\rangle & = &\sum_{d=0}^{p-1}\phi_b(d;\mathbf{k})\sum_{l=-\infty}^{\infty} e^{ilk_1/q+id(2\pi l/p-k_2)} \notag \\
& \times & \delta (v-k_2+2\pi l/p) . 
\end{eqnarray} 
Here the index $b=1,\dots ,p$ labels $p$ QE bands ${\cal E}_b(\mathbf{k})$, where $\mathbf{k}=(k_1,k_2)$ is a Bloch wave vector ranging in the Brillouin zone (BZ)
\begin{equation}\label{BZ}
0\leq k_1 <2\pi q/p ,\ \ \ 0\leq k_2 <2\pi /p,
\end{equation} 
and ${\mathbf V}_b({\mathbf k})=\left\{ \phi_b(d;\mathbf{k})\right\}_{d=0}^{p-1}$ ($b=1,\dots ,p$) are $p$ independent column vectors of coefficients. These are eigenvectors of a $p\times p$ unitary matrix $\hat{M}({\mathbf k})$ with eigenvalues $\exp[-i{\cal E}_b(\mathbf{k})]$. The latter matrix is periodic in $\mathbf{k}$ in a second Brillouin zone (BZ1)
\begin{equation}\label{BZ1}
0\leq k_1 <2\pi,\ \ \ 0\leq k_2 <2\pi /p.
\end{equation}
Let us assume isolated or nondegenerate QE bands, i.e., the $p$ QE eigenvalues at any fixed $\mathbf{k}$ are all different, $\exp[-i{\cal E}_b(\mathbf{k})]\neq \exp[-i{\cal E}_{b'}(\mathbf{k})]$ for $b\neq b'$. Then, the Bloch eigenstates (\ref{qes}) must be periodic in the BZ (\ref{BZ}) up to phase factors that may depend on $b$ and on ${\mathbf k}$. Because of the single valuedness of $|\Psi_{b,{\mathbf k}}\rangle$ in ${\mathbf k}$, the total phase change of $|\Psi_{b,{\mathbf k}}\rangle$ when going around the boundary of the BZ (\ref{BZ}) counterclockwise must be an integer multiple of $2\pi$. This integer, which we denote by $-\sigma_b$, is a Chern integer, a topological characteristic of band $b$. Similarly, the total phase change of ${\mathbf V}_b({\mathbf k})$ when going around the boundary of BZ1 (\ref{BZ1}) counterclockwise must be $2\pi\mu_b$, where $\mu_b$ is a second Chern integer. Assuming ${\mathbf V}_b({\mathbf k})$ to be normalized, one can write:
\begin{eqnarray}\label{mub}
& &\mu_b =\frac{1}{2\pi i}\oint_{\rm BZ1}{\mathbf V}_b^{\dagger}({\mathbf k})\frac{d{\mathbf V}_b({\mathbf k})}{d{\mathbf k}}\cdot d{\mathbf k} \\
&=&\iint_{\rm BZ1}d{\mathbf k}\sum_{b'\neq b}\Im\left\{\frac{{\mathbf V}_b^{\dagger}({\mathbf k})\frac{d\hat{M}^{\dagger}({\mathbf k})}{dk_1}{\mathbf V}_{b'}{\mathbf V}_{b'}^{\dagger}({\mathbf k})\frac{d\hat{M}({\mathbf k})}{dk_2}{\mathbf V}_b}{\pi \left|\exp[-i{\cal E}_{b'}(\mathbf{k})]-\exp[-i{\cal E}_b(\mathbf{k})\right|^2} \right\} , \nonumber
\end{eqnarray}           
where $\Im$ denotes imaginary part and $\hat{M}({\mathbf k})$ is the $p\times p$ matrix defined above; the sum over $b'\neq b$ in the second line of Eq. (\ref{mub}) (following from the first line by use of Stoke's theorem) is Berry's curvature.

The two integers $\sigma_b$ and $\mu_b$ are connected by the Diophantine equation
\begin{equation}\label{de}
p\sigma _{b}+q\mu _{b}=1.
\end{equation}
Derivations of Eq. (\ref{de}) for both static and Floquet systems are based on phase-space translational invariance and can be found in Refs. \cite{daz,dz,d2,d4,ad}; see, in particular, the recent detailed derivations in Refs. \cite{d4,ad}. After $\mu_b$ is calculated from Eq. (\ref{mub}), $\sigma_b$ is determined from Eq. (\ref{de}). Being associated with the eigenvectors ${\mathbf V}_b({\mathbf k})$ of a periodic matrix $\hat{M}({\mathbf k})$, the Chern numbers $\mu_b$ satisfy the sum rule $\sum_{b=1}^{p}\mu_b =0$ \cite{ass,dc}. The latter sum and Eq. (\ref{de}) imply then that
\begin{equation}\label{sr}  
\sum_{b=1}^{p}\sigma_b =1 .
\end{equation}
The Chern number $\sigma_b$ is a topological invariant that can change only at band degeneracies. In consistency with Eqs. (\ref{de}) and (\ref{sr}), the Chern numbers of two neighboring bands $b$ and $b+1$, which degenerate at some parameter values, can change only as follows:
\begin{equation}\label{ccn}  
\sigma_b \rightarrow \sigma_b-lq,\ \ \ \sigma_{b+1} \rightarrow \sigma_{b+1}+lq ,
\end{equation}
where $l$ is some integer; clearly, $\sigma_b+\sigma_{b+1}$ is preserved under the changes (\ref{ccn}). Such changes are topological phase transitions.

\begin{center}
\textbf{IV. TOPOLOGICAL PROPERTIES OF QE BANDS AND PHASE TRANSITIONS IN A SEMICLASSICAL REGIME}
\end{center}

In this section, we study the quantum topological properties of the QE bands of the system (\ref{GH}) and possible topological phase transitions occurring under variation of $E_x$. We shall focus on a semiclassical regime of small rational 
$\hbar_{\rm s}=q/p$ in order to establish a useful correspondence between quantum and classical features. This correspondence can be established in the simplest way in the ``pure" case \cite{dh} of $q=1$, i.e., $\hbar_{\rm s}=1/p$, with sufficiently large $p$. In fact, while the operators (\ref{Uper}) and (\ref{Upa4}) appear to exhibit a $2\pi\times 2\pi$ unit cell of periodicity in the $(u,v)$ phase space, like their classical counterparts (see Appendix B), these operators actually commute only with the translations (\ref{D12}), defining a $2\pi q\times 2\pi$ unit cell. The latter unit cell coincides with the classical one only for $q=1$. We shall therefore assume $\hbar_{\rm s}=1/p$ and also odd $p$, since for even $p$ there may occur band degeneracies for many parameter values. In the next section, we shall explain how our results can be generalized to $q>1$. 

\begin{center}
\textbf{A. Case of $\eta =0$, first $E_x$ sub-interval}
\end{center}

\begin{figure}[tbp]
\includegraphics[width=6.5cm,trim = {1cm 0.2cm 0cm 0cm}]{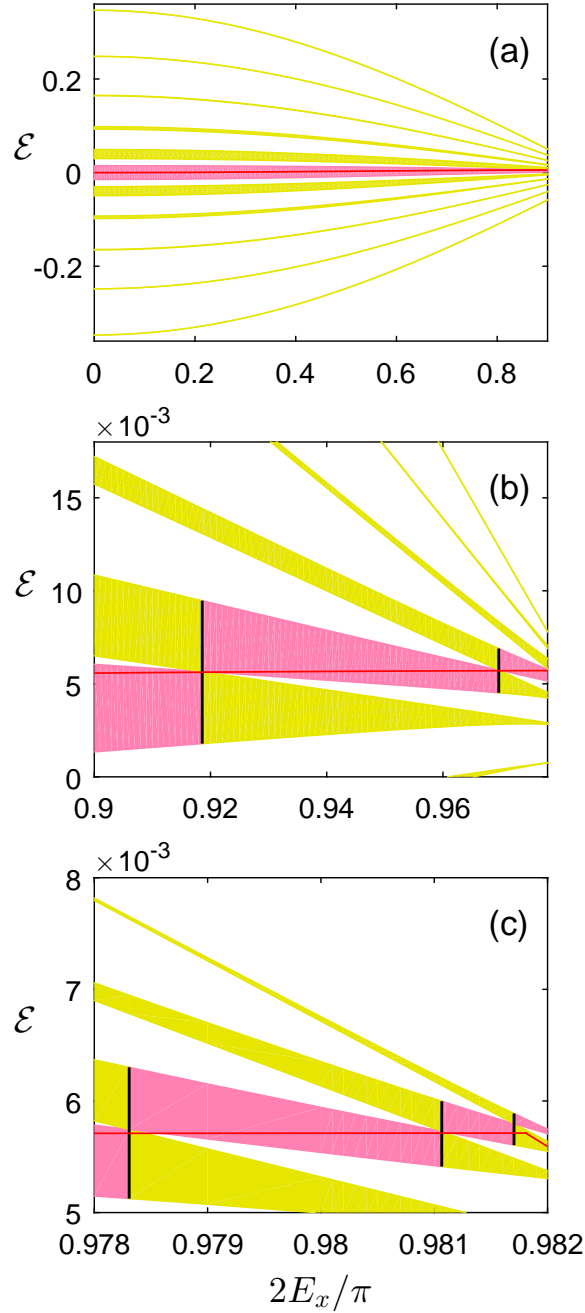}
\caption{(Color online). QE bands ${\cal E}_b(\mathbf{k})$ versus $2E_x/\pi $ for the evolution operator (\ref{Upa4}) ($x_{\rm c}=0$), with $\mathbf{k}$ covering the BZ (\ref{BZ1}), in the case of $\kappa \approx 0.057$ and $\hbar_{\rm s}=1/11$ ($\mu =\kappa/\hbar =0.1$); the yellow (light gray) and pink (gray) regions indicate bands with Chern number $\sigma_b=0$ and $\sigma_b=1$, respectively. Three domains of $2E_x/\pi$ are shown: (a) $0\leq 2E_x/\pi < 0.9$. Here all the $11$ bands (labeled by $b=1,...,11$ in order from bottom to top) have $\sigma_b=0$, except of the central band ($b=6$) with $\sigma_b=1$. (b) $0.9\leq 2E_x/\pi < 0.978$, showing bands $b=5,...,11$. Here two band degeneracies (indicated by vertical black segments) occur between bands $b$ and $b+1$ for $b=6$ (at $2E_x/\pi \approx 0.9186$) and $b=7$ (at $2E_x/\pi \approx 0.9696$). At each degeneracy, the Chern numbers $(\sigma_b,\sigma_{b+1})$ change from $(1,0)$ to $(0,1)$. (c) $0.978\leq 2E_x/\pi < 0.982$, showing bands $b=8,...,11$. Here three band degeneracies occur between bands $b$ and $b+1$ for $b=8$ (at $2E_x/\pi \approx 0.9783$), $b=9$ (at $2E_x/\pi \approx 0.981$), and $b=10$ (at $2E_x/\pi \approx 0.9817$). The Chern numbers change as in (b). The red line in all plots is ${\cal E}=\mu H^{(\rm e)}_{\rm sep}$, where $H^{(\rm e)}_{\rm sep}$ is the classical separatrix energy (\ref{Es}).}      
\label{fig1}
\end{figure}

We consider here the case of $\eta =0$, with the evolution operator given by Eq. (\ref{Upa4}) ($x_{\rm c}=0$). Since this operator is invariant under the two transformations $E_x\rightarrow E_x+\pi,\hat{u}\rightarrow\hat{u}+\pi,\hat{v}\rightarrow\hat{v}+\pi$ and $E_x\rightarrow -E_x,\hat{u}\rightarrow -\hat{u},\hat{v}\rightarrow -\hat{v}$. Thus, the QE spectrum is periodic in $E_x$ with period $\pi$ and has also inversion symmetry around $E_x=0$. Therefore, one can restrict $E_x$ to the interval $0\leq E_x\leq\pi /2$. There are two basically different sub-intervals of this interval to be examined. The first one is $0\leq E_x\leq 0.982\pi /2$. Figure \ref{fig1} shows, in three successive parts of this sub-interval, the QE bands and their Chern numbers $\sigma_b$, calculated as mentioned in Sec. III, for $\hbar_{\rm s}=1/11$ and a small value of $\kappa$. We see that all bands have $\sigma_b=0$ with the exception of one band with $\sigma_b=1$. The latter band is the central one ($b=6$) for $E_x\lesssim 0.918\pi /2$ (Figs. 1(a) and 1(b)). This is consistent with the fact that for sufficiently small $E_x$ and $\kappa$, the effective Hamiltonian (\ref{Hee}) for $\eta =0$ is approximately given by Eq. (\ref{HeQMEta0a}), which is a Harper Hamiltonian \cite{harper}. The Chern numbers for this Hamiltonian were derived by TKNN \cite{tknn} in the context of Bloch electrons in a magnetic field: For $\hbar_{\rm s}=q/p$ (corresponding to $p/q$ magnetic flux quanta per unit cell), the total Chern number of the lowest $b$ bands, $\sigma (b)=\sum_{b'=1}^{b} \sigma_{b'}$, is uniquely determined by the Diophantine equation [corresponding to the sum of Eq. (\ref{de}) over the $b$ bands]
\begin{eqnarray}\label{DiophSum}
p\sigma (b) +q\mu(b)=b,
\end{eqnarray}
where the integer $\mu (b)=\sum_{b'=1}^{b} \mu_{b'}$ satisfies $|\mu (b)|\leq p/2$ \cite{tknn}, except for $b=p/2$ in the case of $p$ even, where there is no gap. For $q=1$ and $p$ odd, Eq. (\ref{DiophSum}) gives $\sigma (b)=0$ for $b\leq (p-1)/2$ and $\sigma (b)=1$ for $b>(p-1)/2$, in accordance with the values above of $\sigma_b$ for $E_x\lesssim 0.918\pi /2$.   

As $E_x$ is increased, however, the effective Hamiltonian (\ref{Hee}) for $\eta =0$ [with Eqs. (\ref{HeQMEta0a}) and (\ref{HeQMEta0b})] starts to deviate considerably from the Harper Hamiltonian. As we see in Figs. 1(b) and 1(c) for $E_x\geq 0.9186\pi /2$, there occur five band degeneracies at some $\mathbf{k}$ between bands $b$ and $b+1$, $b=6,...,10$, at the values of $E_x$ indicated in the caption of Fig. 1. At each degeneracy, the Chern numbers $(\sigma_b,\sigma_{b+1})$ change from $(1,0)$ to $(0,1)$ (topological phase transitions); these changes of $\sigma_b$ correspond to $l=1$ in Eq. (\ref{ccn}) for $q=1$. The value of $\sigma_b=1$ is then transferred from the central band $b=6$ to the highest band, $b=11$.

\begin{figure}[tbp]
\includegraphics[width=7cm,trim = {0.5cm 0.2cm -1cm -0.1cm}]{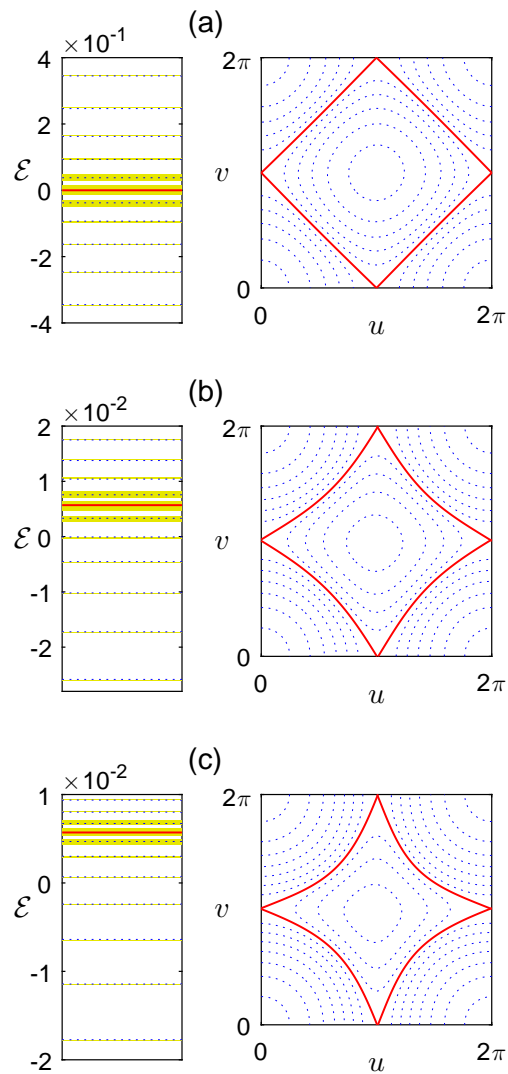}
\caption{(Color online). Left diagrams: QE bands of the evolution operator (\ref{Upa4}) for $x_{\rm c}=0$ [yellow (light gray) regions] and their semiclassical-level approximations (blue dotted lines) for $\kappa \approx 0.057$, $\hbar_{\rm s}=1/11$, and (a) $E_x=0$; (b) $E_x=0.96\pi /2$ (after the first degeneracy in Fig. 1(b)); (c) $E_x=0.975\pi /2$ (after the second degeneracy in Fig. 1(b)). The red solid line is ${\cal E}=\mu H^{(\rm e)}_{\rm sep}$, where $H^{(\rm e)}_{\rm sep}$ is the classical separatrix energy (\ref{Es}). Right diagrams: classical phase spaces of the effective Hamiltonian (\ref{Hec1}) for the same parameter values as above. The red solid line is the separatrix and the blue dotted lines are the classical orbits corresponding to the levels in the left diagrams. As $E_x$ is increased, the area of the separatrix region and the number of orbits inside this region (corresponding to the bands/levels above the separatrix band) decrease.}
\label{fig2}
\end{figure}

To understand these phenomena, we use the approximation of the system by its effective Hamiltonian, Eqs. (\ref{Hee})-(\ref{HeQMEta0b}), which reduces in the classical limit of $\hbar\rightarrow 0$ to
\begin{eqnarray}
H^{(\rm e)}(u,v)&\approx &-2\cos (E_x)[(\cos(u) + \cos (v)] \label{Hec1} \\ 
&- &\frac{\kappa}{2}[\cos(u+v)+\cos(2E_x)\cos(u-v) \notag \\
&- &\sin(2E_x)\sin(u+v)]. \notag
\end{eqnarray}
We plot in Fig. \ref{fig2}, for the same values of ($\kappa$,$\hbar_{\rm s}$) as in Fig. 1 and for some representative values of $E_x$, the closed orbits of the classical system (\ref{Hec1}) satisfying the semiclassical quantization conditions
\begin{equation}\label{scc}
\oint v\, du =2\pi\hbar (l+1/2)=4\pi ^2 \hbar_{\rm s}(l+1/2)
\end{equation}
for integer $l$. There are precisely $p-1=10$ closed orbits satisfying Eq. (\ref{scc}), divided into two groups, see Fig. 2. One group of orbits encircle the stable fixed point at $(u,v)=(0,0)$ while the second group encircles the stable fixed point at $(u,v)=(\pi ,\pi )$. The two groups are separated by the separatrix orbit connecting the unstable fixed points at $(u,v)=(0,\pi )$ and $(u,v)=(\pi ,0)$ [as well as the equivalent points $(u,v)=(2\pi ,\pi )$ and $(u,v)=(\pi ,2\pi )$ in the $2\pi \times 2\pi$ unit cell]. Together with the separatrix, one has $p=11$ orbits. We also plot in Fig. 2 the $p$ exact QE bands and the QE semiclassical levels ${\cal E}=\mu H^{(\rm e)}$ of all these orbits, where $H^{(\rm e)}$ is the classical energy (\ref{Hec1}) under condition (\ref{scc}), except of the separatrix for which no semiclassical condition is imposed. The QE level corresponding to the separatrix is ${\cal E}=\mu H^{(\rm e)}_{\rm sep}$, where $H^{(\rm e)}_{\rm sep}$ is just the classical separatrix energy, calculated at the end of Appendix B:
\begin{equation}\label{Es}
H^{(\rm e)}_{\rm sep} = \kappa \sin^2(E_x) + O(\kappa^3).
\end{equation}
The QE level ${\cal E}=\mu H^{(\rm e)}_{\rm sep}$ is plotted in Fig. 1 versus $E_x$. We see from Figs. 1 and 2 that the separatrix always corresponds to the QE band with $\sigma_b=1$ while the closed orbits correspond to the QE bands with $\sigma_b=0$. This could be expected from the fact that the separatrix orbit is not contractible to a point, since it extends over all a torus, i.e., the $2\pi\times 2\pi$ unit cell of periodicity, in both the $u$ and $v$ directions. On the other hand, all other orbits are localized inside the unit cell and are therefore contractible to a point. This topological difference between the separatrix and the other orbits is expressed by the nonzero value of $\sigma_b=1$ for the separatrix band, in contrast with $\sigma_b=0$ for the other bands. We see from Fig. 1 that $\mu H^{(\rm e)}_{\rm sep}$ varies over the QE spectrum from the central band $b=(p+1)/2=6$ to the highest band $b=p=11$. Also, when $\mu H^{(\rm e)}_{\rm sep}$ is in between bands $b$ and $b+1$, $b\geq 6$, it is always close to the degeneracy point between these two bands. All this explains the $(p-1)/2=5$ topological phase transitions occurring in the $E_x$ domains of Fig. 1.

\begin{center}
\textbf{B. Case of $\eta =0$, second $E_x$ sub-interval}
\end{center}      

\begin{figure}[tbp]
\includegraphics[width=6.25cm,trim = {1cm 0.2cm -1cm 0cm}]{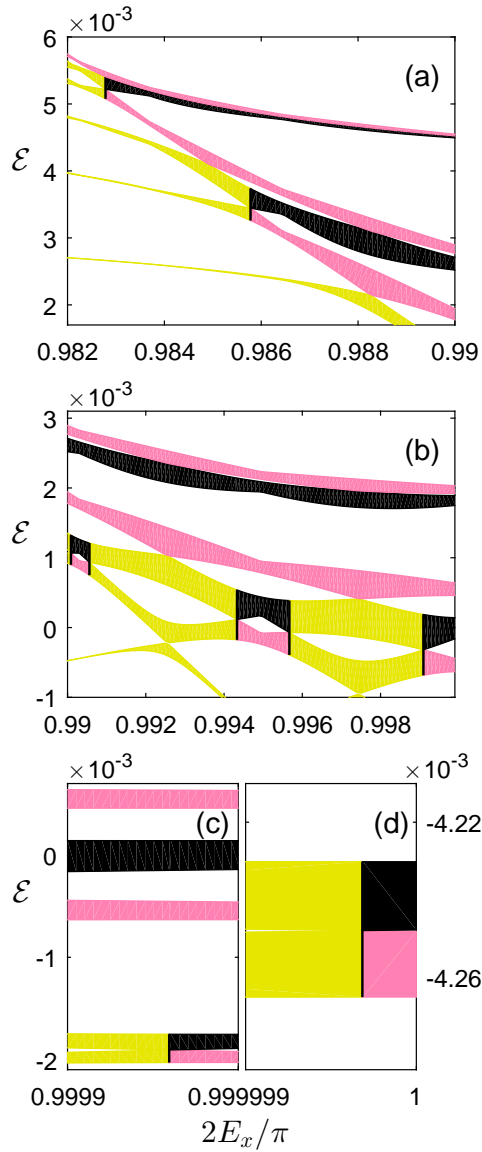}
\caption{(Color online). Continuation of Fig. 1 to the sub-interval $0.982< 2E_x/\pi \leq 1$; here there appear also bands with Chern number $\sigma_b=-1$ (black regions): (a) $0.982< 2E_x/\pi \leq 0.99$, showing bands $b=6,...,11$. Band degeneracies occur at $2E_x/\pi \approx 0.9828$ (bands $b=9,10$) and at $2E_x/\pi \approx 0.9858$ (bands $b=7,18$). (b) $0.99< 2E_x/\pi \leq 0.9999$, showing bands $b=4,...,9$. Degeneracies between bands $b=5,6$ occur at $2E_x/\pi \approx 0.9901,0.9906,0.9943,0.9957,0.9991$. (c) $0.9999< 2E_x/\pi \leq 0.999999$, showing bands $b=3,...,7$. A degeneracy between bands $b=3,4$ occurs at $2E_x/\pi \approx 0.99996$. (d) $0.999999< 2E_x/\pi \leq 1$, showing bands $b=1,2$ which degenerate  at $2E_x/\pi \approx 0.9999997$.}  
\label{fig3}
\end{figure}

We now examine the continuation of the QE bands above to the sub-interval $0.982< 2E_x/\pi \leq 1$, shown again in three parts in Fig. \ref{fig3}. This sub-interval is basically different from that considered in Sec. IV A (Fig. 1) in that a significant change in the classical phase-space structure occurs for $2E_x/\pi \gtrsim 0.982$; see details in Appendix B, third section. In particular, the form of the separatrix completely changes, compare the separatrices in Figs. 2 (right diagrams), 7(a), and 7(b) with those in Figs. 7(g) and 7(h). The metamorphosis in the separatrix shape indeed occurs 
near $2E_x/\pi \approx 0.982$, see Figs. 7(c), 7(d), 7(e), and 7(f). 

Figure 3(a) shows that, due to a degeneracy of bands $b=9$ and $b=10$ at $2E_x/\pi \approx 0.9828$, $(\sigma_9,\sigma_{10})$ change from $(0,0)$ to $(1,-1)$. As $E_x$ is increased, band $b=10$ ($\sigma_{10}=-1$) approaches band $b=11$ ($\sigma_{11}=1$) and these two bands then form a 2-band cluster with total Chern number $\sigma_{10,11}=0$. This cluster persists up to the final value of $2E_x\pi =1$. Similarly, after the degeneracy of bands $b=7$ and $b=8$ at $2E_x/\pi \approx 0.9858$, the band $b=8$ ($\sigma_{8}=-1$) approaches band $b=9$ ($\sigma_{9}=1$), forming a 2-band cluster with $\sigma_{8,9}=0$ (see Figs. 3(a) and 3(b)), which again persists up to $2E_x\pi =1$. Other two 2-band clusters, corresponding to the two pairs of bands $(b,b+1)$ ($b=1,3$) with $(\sigma_b=1,\sigma_{b+1}=-1)$, are formed for $2E_x\pi$ very close to $1$ (see Figs. 3(c) and 3(d)). Finally, consider the three bands $b=5,6,7$. After the last (third) degeneracy in Fig. 3(b), at $2E_x/\pi \approx 0.9991$, $(\sigma_5,\sigma_6)$ change from $(0,0)$ to $(1,-1)$. As shown in Figs. 3(b) and 3(c), these two bands remain relatively close to band $b=7$ with $\sigma_7=1$ up to $2E_x\pi =1$. The three bands $b=5,6,7$ then form a 3-band cluster with total Chern number $\sigma_{5,6,7}=1$.

The classical-quantum correspondence in the case of band clusters is clearly exhibited using the concept of band Husimi distribution (BHD) \cite{drf}. The BHD of a cluster of $N$ adjacent bands, $b,b+1,...,b+N-1$, is defined in the $(u,v)$ phase space by
\begin{equation}\label{BHD}
P_{b,N}(u,v)=\frac{1}{N|{\rm BZ}|}\sum_{b'=b}^{b+N-1}\int d{\mathbf k}\left|\left\langle u,v|\Psi_{b',\mathbf{k}}\right\rangle \right|^2 ,
\end{equation}
where $|{\rm BZ}|=4\pi^2q/p^2$ is the area of the BZ (\ref{BZ}) and $\left\langle u,v|\Psi_{b,\mathbf{k}}\right\rangle$ is the coherent-state representation of the QE band eigenstates (\ref{qes}).

Examples of BHDs for some of the band clusters considered above are shown in Fig. 4 for the maximal value of $2E_x/\pi =1$, where we plot also the classical separatrix which is clearly very different from the separatrix in the first $E_x$ sub-interval in Fig. 2. We see from Fig. 4 that the BHD for the 2-band cluster of $b=10,11$ (Fig. 4(a)) and the 2-band cluster of $b=1,2$ (Fig. 4(c)) is concentrated on localized states inside or outside the separatrix regions. This is consistent with the fact that the two 2-band clusters are both associated with a total Chern number equal to zero, see above. On the other hand, the BHD for the 3-band cluster of $b=5,6,7$ (Fig. 4(b)), is concentrated on the separatrix, mainly on the four hyperbolic fixed points. This is consistent with the nonzero total Chern number $\sigma_{5,6,7}=1$ for this cluster. The formation of band clusters with total Chern numbers as above, for $2E_x/\pi$ approaching 1, will be understood in the next section, IV C.     

\begin{figure}[tbp]
\includegraphics[width=6.5cm,trim = {1cm 0.2cm 0cm 0cm}]{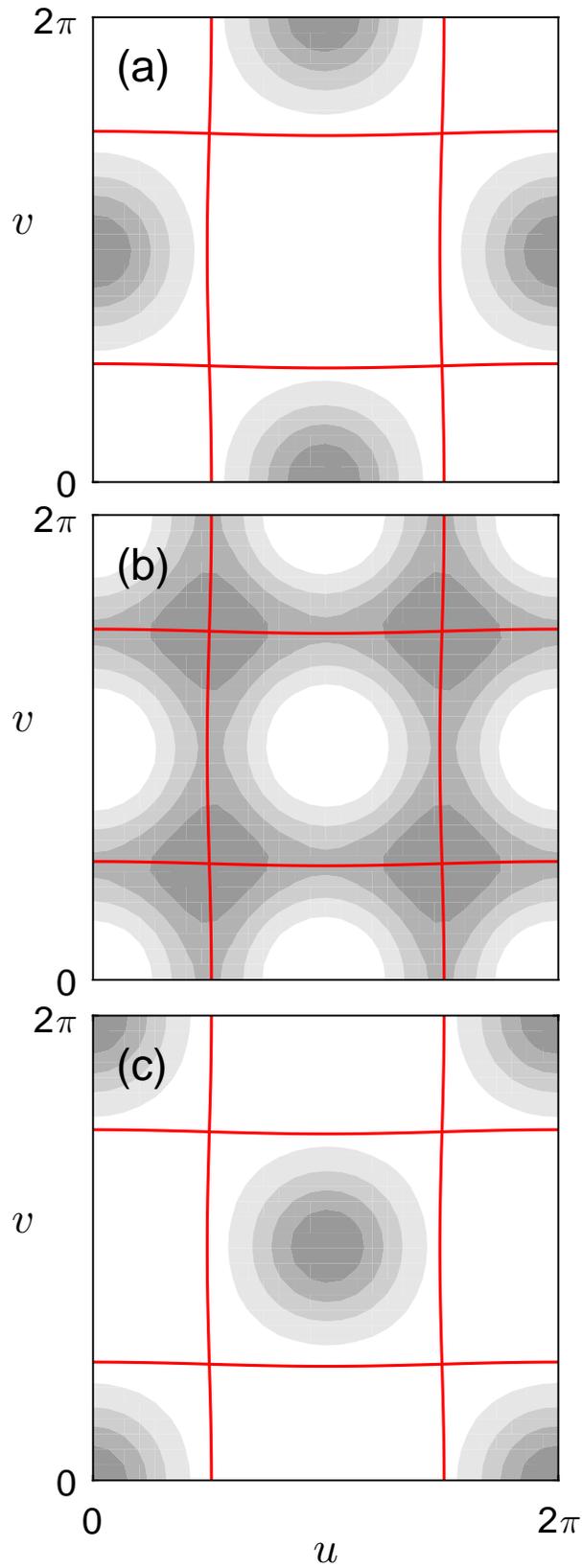}
\caption{(Color online). Density plots of BHDs (\ref{BHD}) for some band clusters considered in the text (darker regions correspond to higher values of the BHD) at the maximal value of $2E_x/\pi =1$, see also Fig. 3: (a) 2-band cluster of $b=10,11$; (b) 3-band cluster of $b=5,6,7$; (c) 2-band cluster of $b=1,2$. The red solid lines in all plots define the classical separatrix for $2E_x/\pi =1$, see also Fig. 7(h).}  
\label{fig4a}
\end{figure}
 
\begin{center}
\textbf{C. Case of $\eta \neq 0$}
\end{center}

We now consider the case of electric fields having a nonzero component $E_y$ satisfying the resonance conditions (\ref{rc}) with $r={\rm lcm}(4,\ell )>8$. In this case, the lowest terms of the effective Hamiltonian are given by Eqs. (\ref{HeQMEtaNon0a}) and (\ref{HeQMEtaNon0b}). These terms coincide, up to a constant factor, with those for $\eta =0$ and $E_x=\pi /2$ [see Eqs. (\ref{HeQMEta0a}) and (\ref{HeQMEta0b})], a case considered in Sec. IV B.

Figure 5 shows the QE bands of the exact evolution operator (\ref{Uper}) as functions of $2E_x/\pi$ for $\eta /(2\pi )=2/3$ ($r=12$) and several values of $\mu =\kappa/\hbar$ and $\hbar_{\mathrm{s}}$.  As expected from Eqs. (\ref{HeQMEtaNon0a}) and (\ref{HeQMEtaNon0b}), that are independent of $E_x$, the bands exhibit approximately this independence. These bands also form three clusters in all the cases of Fig. 5. We now explain the reason for the formation of these clusters, as well as those for $\eta =0$ and $E_x$ close to $\pi /2$ in Sec. IV B. 

\begin{figure}[tbp]
\includegraphics[width=6.5cm,trim = {-0.5cm 0.2cm -1cm 0.5cm}]{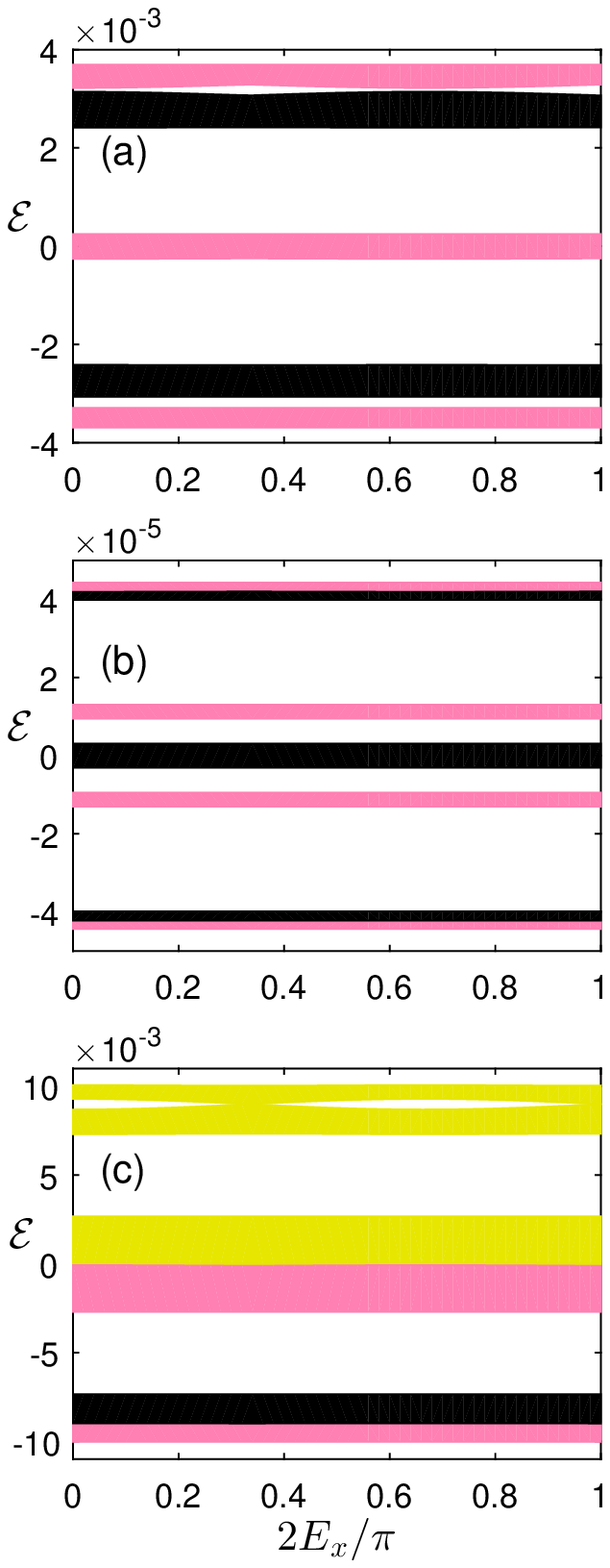}
\caption{(Color online). QE bands as functions of $2E_x/\pi$ for $\eta /(2\pi)=2/3$ and: (a) $\mu=0.04$, $\hbar_{\mathrm{s}}=1/5$; (b) $\mu=0.005$, $\hbar_{\mathrm{s}}=1/7$; (c) $\mu =0.07$, $\hbar_{\mathrm{s}}=1/6$. The Chern numbers of the bands are indicated by the colors as in Figs. 1 and 3. In all cases, one can see three band clusters, with the central one consisting of one band [in (a)], three bands [in (b)], and two bands [in (c)].}  
\label{fig5}
\end{figure} 

To first order in the expansion (\ref{Hee}), with Eqs. (\ref{HeQMEtaNon0a}) and (\ref{HeQMEtaNon0b}), the effective Hamiltonian is
\begin{equation}\label{He1}
\hat{H}^{(\rm e)}(\hat{u},\hat{v})\approx -\frac{\epsilon r}{8\cos (\eta)}\big[ \cos(\hat{u}')+\cos(\hat{v}')\big],
\end{equation}
after defining the new quantum variables $\hat{u}'=\hat{u}+\hat{v}$ and $\hat{v}'=\hat{u}-\hat{v}$. These variables satisfy $[\hat{u}',\hat{v}']=2\pi i \hbar_{\mathrm{s}}^\prime$, where $\hbar_{\mathrm{s}}^{\prime}=2\hbar_{\mathrm{s}}$.  The meaning of the latter relation is as follows. The Hamiltonian (\ref{He1}) is $2\pi$-periodic in both variables $(\hat{u}',\hat{v}')$. This periodicity corresponds, in the original variables $(\hat{u},\hat{v})$, to a periodicity with a square unit cell defined by the vectors $\pi (1,1)$ and $\pi (-1,1)$. The area of this unit cell is $2\pi^2$, half of the unit-cell area $4\pi^2$ of the general Hamiltonian (\ref{Hee}). Thus, the effective Planck constant $\hbar_{\mathrm{s}}^{\prime}$ is twice the ordinary one $\hbar_{\mathrm{s}}$. The approximate periodicity with the new unit cell above is evident from Fig. 4. Also, the classical separatrix in these figures corresponds to the ``energy surface" $H^{(\rm e)}(u,v)=0$ \cite{bhd}, where     
$H^{(\rm e)}(u,v)$ is the classical limit $\hbar_{\mathrm{s}}\rightarrow 0$ [with $\epsilon\rightarrow \kappa /2$ in Eq. (\ref{eps})] of the Hamiltonian (\ref{He1}). In the original variables:
\begin{equation}\label{He1c}
H^{(\rm e)}(u,v)\approx -\frac{\kappa r}{8\cos (\eta)}\cos(u)\cos(v), 
\end{equation} 
so that $H^{(\rm e)}(u,v)=0$ corresponds to the four lines $u=\pi /2,3\pi /2$ and $v=\pi /2,3\pi /2$. These lines define indeed a first approximation of the separatrix in Fig. 4.

Now, the quantum Hamiltonian (\ref{He1}) is the Harper one \cite{harper}, whose energy bands have Chern numbers uniquely determined by the TKNN Eq. (\ref{DiophSum}). Specifically, in the case of $\hbar_{\mathrm{s}}=1/p$ with $p$ odd, one has $\hbar_{\mathrm{s}}^{\prime}=2\hbar_{\mathrm{s}}=q'/p'$ with $q'=2$ and $p'=p$. Using Eq. (\ref{DiophSum}) for the primed quantities, we get for $q'=2$ that $\sigma_b^{\prime} = \sigma^{\prime}(b)-\sigma^{\prime}(b-1)$ can take only the values $\sigma_{b}^{\prime}=\pm 1$. These are, indeed, the values of the Chern numbers for $2E_x/\pi =1$ in Fig. 3.

Next, it is known that for $q'>1$ (in particular, $q'=2$) the energy bands of Harper-like Hamiltonians generally form band clusters \cite{dh,mw}. Consider the case of $q'/p'$ very close to $\bar{q}/\bar{p}$, $p'\gg \bar{p}$. Let $\bar{\sigma}_b$ be the Chern numbers of the $\bar{p}$ bands for $\hbar_{\mathrm{s}}=\bar{q}/\bar{p}$. From the results in Ref. \cite{mw}, a band $b$ for $\hbar_{\mathrm{s}}=\bar{q}/\bar{p}$ will ``split" into a cluster of $N_b$ bands for $\hbar_{\mathrm{s}}=q'/p'$, where 
\begin{equation}\label{nb}
N_b=p'\bar{\sigma}_b+q'\bar{\mu}_b ,
\end{equation}
and $\bar{\mu}_b$ is the integer satisfying the Diophantine equation $\bar{p}\bar{\sigma}_b+\bar{q}\bar{\mu}_b=1$. Also, the total Chern number of the cluster is equal to $\bar{\sigma}_b$. 

Let us apply these results to the case above of $\hbar_{\mathrm{s}}^{\prime}=2/p'$, choosing $\bar{q}/\bar{p}=1/\bar{p}$, where $p'$ and $\bar{p}$ are odd integers. For $\hbar_{\mathrm{s}}=1/\bar{p}$, the Chern numbers of the $\bar{p}$ bands are zero except that of the central band, $b=(\bar{p}+1)/2$, with $\bar{\sigma}_b=1$ (see first paragraph of Sec. IV A). Then, $\bar{\mu}_b=1$ except for $b=(\bar{p}+1)/2$, with $\bar{\mu}_b=1-\bar{p}$. This implies, from Eq. (\ref{nb}) with $q'=2$, that each $\bar{\sigma}_b=0$-band will split into a cluster of two bands for $\hbar_{\mathrm{s}}=\hbar_{\mathrm{s}}^{\prime}$; these bands must have Chern numbers $\sigma_b^{\prime}=1,-1$ since these are the only possible values of $\sigma_b^{\prime}$ for $q'=2$ (see above) and since the total Chern number of the cluster must be equal to $\bar{\sigma}_b=0$. Similarly, we find that the central band $b=(\bar{p}+1)/2$, with $\bar{\sigma}_b=1$, will split into a cluster of $N_b=p'+2-2\bar{p}$ with total Chern number equal to $1$; thus, in this cluster there must be $(p'+3)/2-\bar{p}$ bands with $\sigma_b^{\prime}=1$ and $(p'+1)/2-\bar{p}$ bands with $\sigma_b^{\prime}=-1$. For example, in the case of $p=p'=11$ with $\hbar_{\mathrm{s}}^{\prime}=2/11$, we have $\bar{p}=5$, so that the central cluster will consist of $N_b=3$ bands, two with $\sigma_b^{\prime}=1$ and one with $\sigma_b^{\prime}=-1$. This is in perfect accordance with the numerical observations in Sec. IV B. In the same way, one can verify that the band clusters and Chern numbers for $p=5$ and $p=7$ (both with $\bar{p}=3$ band clusters), are as shown in Figs. 5(a) and 5(b).

Finally, consider the case of $p$ even, $p=2\bar{p}$, and assume for simplicity that $\bar{p}$ is odd (as in the example of Fig. 5(c), with $p=6$ and $\bar{p}=3$). Then, $\hbar_{\mathrm{s}}^{\prime}=1/\bar{p}$, so that the spectrum of the Harper Hamiltonian (\ref{He1}) will consist of $\bar{p}$ bands, with the central band $b=(\bar{p}+1)/2$ having $\bar{\sigma}_b=1$ and $\bar{\mu}_b=1-\bar{p}$; all other bands have $\bar{\sigma}_b=0$ and $\bar{\mu}_b=1$. This and Eq. (\ref{nb}) (with $p'=p$ and $q'=2$) imply that each band $b$ will correspond, for $\hbar_{\mathrm{s}}=1/p$, to a cluster of two bands whose total Chern number is equal to $\bar{\sigma}_b$, in accordance with Fig. 5(c). However, the Chern numbers of the two bands cannot be determined in this case from general considerations. In fact, they may assume general values summing to $\bar{\sigma}_b$, as shown in Fig. 5(c).   
   
\begin{center}
\textbf{V. CASE OF $\eta =0$ FOR $q>1$ AND THE IRRATIONAL LIMIT}
\end{center}

In Sec. IV A and IV B, we considered the case of $\eta =0$ for $\hbar_{\mathrm{s}}=1/p$ ($q=1$) and we have shown that topological phase transitions occur as $E_x$ is varied. Here we shall consider the case of $\eta =0$ for $\hbar_{\mathrm{s}}=q/p$ ($q>1$), including the irrational limit of $q,\ p\rightarrow \infty$. We shall examine how the topological phase transitions occurring for $\hbar_{\mathrm{s}}=1/p$ are manifested in the case of rational or irrational values of 
$\hbar_{\mathrm{s}}$ that are are sufficiently close to $1/p$.

We start from the Diophantine equation (\ref{DiophSum}) for the total Chern number $\sigma (b)$ of the lowest $b$ bands and divide this equation by $p$. We then take the limit of $q,\ p\rightarrow \infty$ of irrational $\hbar_{\mathrm{s}}$, choosing the lowest $b$ bands ($b\rightarrow \infty$) to be those below some fixed gap. Denoting $b/p$ in this limit by $\zeta$, $0<\zeta<1$, we obtain  
\begin{eqnarray}\label{DiophSumIrr}
\sigma + \hbar_{\mathrm{s}} \mu = \zeta,    
\end{eqnarray} 
where $\sigma =\sigma (b\rightarrow \infty)$ and $\mu =\mu (b\rightarrow \infty)$ are integers, so that $\zeta$ must be  irrational like $\hbar_{\mathrm{s}}$. Clearly, Eq. (\ref{DiophSumIrr}) can have only one solution $(\sigma ,\mu)$ at fixed $\zeta$ (see note \cite{note1}), where $\zeta$ characterizes the fixed gap. This means that if this gap closes and reopens under variation of some parameters, the value of $\sigma$ will {\em not} change, i.e., {\em no} topological phase transitions can occur for irrational $\hbar_{\mathrm{s}}$. It is plausible to assume that this will hold to some extent also for rational values of $\hbar_{\mathrm{s}}=q/p$ ($q>1$) that are sufficiently close to some irrational $\hbar_{\mathrm{s}}$. We show below that this is indeed the case already for $q=2$ and we explain how this is compatible with the topological phase transitions occurring for $\hbar_{\mathrm{s}}=1/p$.     
    
Consider the example of the irrational value $\hbar_{\mathrm{s}}=1/(5+\varrho)$ with $\varrho=(\sqrt{5}-1)/2$ being the inverse of the golden mean. Expressing $\varrho$ as a continued fraction, one gets the series of rational approximants of $\hbar_{\mathrm{s}}$: $1/5$, $2/11$, $3/17$,.... Figure 6(a) shows QE bands for $\hbar_{\mathrm{s}}=1/5$ and $\mu=0.1$ in an interval of $2E_x/\pi$ where the first band degeneracy occurs, between bands $b=3$ and $b=4$ at $2E_x/\pi\approx 0.9135$. As a consequence of this degeneracy, the Chern number of band $b=3$ ($b=4$) changes from $1$ ($0$) to $0$ ($1$).

Figure 6(b) shows QE bands for $\hbar_{\mathrm{s}}=2/11$ and $\mu=0.1$ in the same interval of $2E_x/\pi$ and in the same QE range as in Fig. 6(a). There are $11$ bands which ``split" from the $5$ bands for $\hbar_{\mathrm{s}}=1/5$. The $11$ bands have alternating Chern numbers $\sigma_b=(-1)^{b+1}$, $b=1,...,11$. Now, one should observe the following. For small $2E_x/\pi\approx 0.84$, the $11$ bands can be grouped into five clusters, $C_1$, $C_2$, $C_3$, $C_4$, and $C_5$, formed by bands $b=(1,2)$, $b=(3,4)$, $b=(5,6,7)$, $b=(8,9)$, and $b=(10,11)$, respectively. The cluster $C_j$, $j=1,...,5$, splits from band $j$ for $\hbar_{\mathrm{s}}=1/5$ and its total Chern number is equal to the Chern number of this band $j$. Next, one can see from Fig. 6(b) that as $2E_x/\pi$ increases, band $b=7$ in cluster $C_3$ approaches cluster $C_4$ and is then naturally associated to $C_4$ rather than to $C_3$. Thus, the total Chern number of $C_3$ ($C_4$) changes from $1$ ($0$) to $0$ ($1$), in accordance with the changes above for the corresponding bands in the case of $\hbar_{\mathrm{s}}=1/5$. Unlike the latter case, however, the changes of the Chern numbers of clusters occur not due to degeneracies between clusters but due to the transfer of a band from one cluster to a neighboring one.

\begin{figure}[tbp]
\includegraphics[width=6.5cm,trim = {-0.5cm 0.2cm -1cm 0.5cm}]{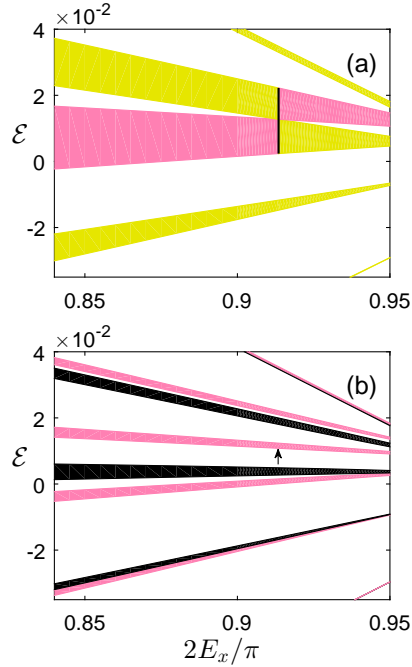}
\caption{(Color online). QE bands as functions of $2E_x/\pi$ for $\mu =0.1$ and: (a) $\hbar_{\mathrm{s}}=1/5$; (b)
$\hbar_{\mathrm{s}}=2/11$. The Chern numbers of the bands are indicated by the colors as in Figs. 1 and 3. One can see in (b) five band clusters (specified in the text), where band $b=7$ (indicated by a small vertical arrow) is associated with the third cluster for small $2E_x/\pi$ but it is more naturally associated with the fourth cluster for sufficiently large $2E_x/\pi$.}  
\label{fig6}
\end{figure} 

We found that the latter scenario holds for almost all $E_x$ except of small $E_x$ intervals $[E_x^{(1)},E_x^{(2)}]$ where the Chern numbers of two clusters change due to a degeneracy between them at $E_x=E_x^{(1)}$ but this change is canceled due to a degeneracy at $E_x=E_x^{(2)}$. Also, degeneracies between bands in the same cluster can occur but the changes in the band Chern numbers (which keep fixed the total Chern number of the cluster) usually occur only in small $E_x$ intervals. These changes within a cluster can be again explained, in the framework of the next rational approximant ($\hbar_{\mathrm{s}}=3/17$), by the transfer of a band from one cluster to another without any degeneracy between the two clusters. In this way, one can understand the absence of degeneracies and topological phase transitions in the irrational limit, as shown above. 

\begin{center}
\textbf{VI. SUMMARY AND CONCLUSIONS}
\end{center}

In this paper, we have performed a first study of the topological properties of a class of Floquet systems basically different from other such systems considered in previous works. This class of systems are described by the quantum Hamiltonian (\ref{GH}) with the time-periodic kicking potential (\ref{Vk}) and under the resonance conditions (\ref{rc}). The unique feature of these systems is the presence of a Hall effect due to perpendicular magnetic ($\mathbf{B}$) and electric ($\mathbf{E}$) fields, where $\mathbf{B}$ is perpendicular also to the kicking ($x$) direction and $\mathbf{E}$ is in the $(x,y)$ plane. These systems are a Floquet version of the static systems in Refs. \cite{km1,km2}, in which the time-periodic 1D potential in Eq. (\ref{GH}) is replaced by a time-independent 2D periodic potential. The finite value of $E$ in these static systems replaces the infinitesimal one in the linear-response theory used by TKNN \cite{tknn}. However, while the quantum dynamics of these systems was extensively investigated in Refs. \cite{km1,km2}, their topological properties were not studied.   

In the case that $\mathbf{E}$ has a nonzero component $E_y$ [$\eta\neq 0$ in Eq. (\ref{rc})], there is a nonzero component of the Hall velocity in the kicking direction. It is then known \cite{bhd,dk}] that the latter component causes, for small kicking strength $\kappa$, a significant reduction of classical and quantum dynamical rates relative to the case of $E_y=0$ or $\eta =0$. This is because for $\eta\neq 0$ the leading term in the effective-Hamiltonian expansion (\ref{Hee}) is of order $\epsilon$ or $\kappa$ while for $\eta =0$ the leading term is of order $O(1)$ (see Sec. II C).

We have shown that this dynamical difference between the cases of $\eta =0$ and $\eta\neq 0$ for small $\kappa$ has a clear manifestation in the topological properties of the system in the two cases. For $\eta =0$, the dependence of the leading term (\ref{HeQMEta0a}) on $E_x$ causes the QE bands to be $E_x$ dependent and thus to exhibit topological phase transitions as $E_x$ is varied, see Figs. 1 and 3. As explained at the end of Sec. II B, this case of $\eta =0$ ($E_y=0$) is essentially the same as that of the system without electric field ($\mathbf{E}=0$) for all $x_{{\rm c}}$, whose quntum-dynamical properties were studied in Ref. \cite{dd1}. Thus, the topological properties and phase transitions for $\eta =0$ as $E_x$ is varied are practically those of the $\mathbf{E}=0$ system as functions of $x_{{\rm c}}$. 

On the other hand, the system for $\eta\neq 0$ and $r={\rm lcm}(4,\ell )>8$ [with $\ell$ defined by Eq. (\ref{rc})] is topologically different from any Floquet system considered until now. This is because in this case the leading term (\ref{HeQMEtaNon0b}) does not depend on $E_x$, so that for sufficiently small $\kappa$ the QE bands are almost independent on $E_x$ and no topological phase transitions occur by varying $E_x$, see Fig. 5. The QE bands are then associated with universal values $\sigma_b=\pm 1$ of the Chern numbers (see Figs. 5(a) and 5(b)), as determined from the Diophantine equations. These Chern numbers are also exhibited in the case of $\eta =0$ for $E_x=\pi /2$ since in this case the leading terms (\ref{HeQMEta0a}) and (\ref{HeQMEta0b}) coincide, up to a constant factor, with the terms (\ref{HeQMEtaNon0a}) and (\ref{HeQMEtaNon0b}).

Since the system (\ref{GH}) with the potential (\ref{Vk}) is essentially equivalent to a modulated kicked harmonic oscillator [see Eq. (\ref{eKHOs}) with Eq. (\ref{xct})], this system may be experimentally realizable as it was done for the ordinary quantum kicked harmonic oscillator using either atom-optics methods with Bose-Einstein condensates \cite{dmcw} or paraxial-optics methods with light beams \cite{lgwrt}.

In future works, we plan to to study the topological phase transitions in our system as $\kappa$ is increased. These transitions are expected to occur also in the case of $\eta\neq 0$. We also plan to investigate possible quantum-transport meanings of the Chern numbers in different cases.    

\renewcommand{\theequation}{A\arabic{equation}}
\setcounter{equation}{0}
\begin{center}
\textbf{APPENDIX A}
\end{center}

We present here the main lines of the derivation of results in Sec. II; see also works \cite{dd1,dk}.

{\it Basic evolution operator}. Let us write $(\hat{u}^2+\hat{v}^2)/2=\hbar (\hat{a}^{\dagger}\hat{a}+1/2)$ in Eq. (\ref{eKHOs}), where $\hat{a}=(\hat{v}-i\hat{u})/\sqrt{2\hbar}$. In the case of the potential (\ref{Vk}), the one-period evolution operator from $t=-0$ to $t=T-0$ is given by
\begin{equation}\label{U1}
\hat{U}=\hat{U}_{\gamma}\hat{U}'_{\hat{x}_{\mathrm{c}},\hat{y}_{\mathrm{c}}}\exp\left[i\mu \cos (\hat{x}_{\mathrm{c}}-E_x
+\hat{v}) \right],
\end{equation}
where $\hat{U}_{\gamma}=\exp\left[-i\gamma(\hat{a}^{\dagger}a+1/2) \right]$ ($\gamma =\omega T=T$), $\hat{U}'_{\hat{x}_{\mathrm{c}},\hat{y}_{\mathrm{c}}}=\exp [-i(E_xT\hat{x}_{\mathrm{c}}+\eta\hat{y}_{\mathrm{c}})/\hbar ]$ ($\eta =E_yT$), and $\mu =\kappa /\hbar$. The operator $\hat{U}_{\gamma}$ is a rotation by angle $\gamma$ in the $(u,v)$ phase plane: $\hat{U}_{\gamma}f(\hat{a}^{\dagger},\hat{a})\hat{U}^{-1}_{\gamma}=f(\hat{a}^{\dagger}e^{-i\gamma},\hat{a}e^{i\gamma})$, for arbitrary function $f(\hat{a}^{\dagger},\hat{a})$ \cite{amp}. In the case assumed in this paper, i.e., $\gamma =\pi /2$, this is a clockwise rotation by $\pi /2$: $v\rightarrow u\rightarrow -v\rightarrow -u\rightarrow v$; thus, one has $\hat{U}_{\gamma}^4=-1$. Also, from $[\hat{x}_{\mathrm{c}},\hat{y}_{\mathrm{c}}]=i\hbar$, one has $\hat{y}_{\mathrm{c}}=-i\hbar d/d{x}_{\mathrm{c}}$, so that $\exp (-i\eta\hat{y}_{\mathrm{c}}/\hbar )=\exp(-\eta d/d{x}_{\mathrm{c}})$, a translation of $\hat{x}_{\mathrm{c}}$ by $-\eta$. Using then all the other commutation relations (\ref{cr}) and restricting our attention to wave functions depending only on the $(u,v)$ degree of freedom and not on $(y_{\mathrm{c}},x_{\mathrm{c}})$, we find that the basic evolution operator $\hat{U}^r$ [$r=\mathrm{lcm}(n=4,\ell )$] is given by the right-hand side of Eq. (\ref{Uper}), after omitting nonrelevant terms including $\hat{U}'_{\hat{x}_{\mathrm{c}},\hat{y}_{\mathrm{c}}}$ and some constant phase factors. 

{\it Effective Hamiltonian}. The effective Hamiltonian $\hat{H}^{(\rm e)}$, giving the basic evolution operator (\ref{Uper}) as $\hat{U}^r=\exp [-i\mu \hat{H}^{(\rm e)}]$, is calculated as follows. Assuming $x_{\mathrm{c}}=0$, as explained at the end of Sec. II B, let 
\begin{eqnarray}\label{Vt}
& & \hat{O}_j =  i\mu\cos (-E_x-j\eta
+\hat{v}_j) \notag \\  & = & i\frac{\mu}{2}\{ \exp [-i(E_x+j\eta )]\exp(i\hat{v}_j)+{\rm c.c.} \} .
\end{eqnarray}
Now, given two operators $\hat{A}$ and $\hat{B}$, one has \cite{rmw}
\begin{eqnarray}\label{id}
& & \exp(\hat{A}) \exp(\hat{B}) = \exp\left(\hat{A} + \hat{B}+\frac{1}{2}[\hat{A},\hat{B}] \right. \notag \\
& & \left. +\frac{1}{12}[\hat{A},[\hat{A},\hat{B}]]+
\frac{1}{12}[[\hat{A},\hat{B}],\hat{B}]+\dots \right) , 
\end{eqnarray} 
involving a series of repeated commutators on the right-hand side. Equation (\ref{id}) can be applied to derive systematically an expansion for $\hat{H}^{(\rm e)}$. From the definition of $\hat{v}_j$ after Eq. (\ref{Uper}), we see that $[\hat{v}_j,\hat{v}_{j+1}]=-i\hbar$. Therefore, the commutator $[\hat{O}_j,\hat{O}_{j+1}]$ of two adjacent operators (\ref{Vt}) will be a linear combination of commutators of the form
\begin{equation}\label{c1}
\left[e^{ig_1\hat{v}_j},e^{ig_2\hat{v}_{j+1}}\right]=2i\sin (g_1g_2\pi\hbar_{\rm s}) e^{i(g_1\hat{v}_j+g_2\hat{v}_{j+1})}
\end{equation}        
for $g_1,g_2=\pm 1$, after using Eq. (\ref{id}) with $[\hat{v}_j,\hat{v}_{j+1}]=-i\hbar$ and denoting $\hbar_{\rm s}=\hbar/(2\pi )$. More generally, for integers $g_1$, $g_2$, $g_3$, $g_4$,
\begin{eqnarray}\label{c2}
& & \left[e^{i(g_1\hat{u}+g_2\hat{v})},e^{i(g_3\hat{u}+g_4\hat{v})}\right] \\ & & = 2i\sin [(g_2g_3-g_1g_4)\pi\hbar_{\rm s}] e^{i(g_1+g_3)\hat{u}+i(g_2+g_4)\hat{v}}. \notag
\end{eqnarray}
We also note that for non-zero integer $a$ one has
\begin{equation}\label{id1}
\sin (a\pi\hbar_{\rm s})=J(a;\hbar_{\rm s})\sin (\pi\hbar_{\rm s}),
\end{equation}
where the function $J(a;\hbar_{\rm s})$ does not vanish for integer $\hbar_{\rm s}$. It is then easy to see that Eqs. (\ref{id})-(\ref{id1}) imply the expansion (\ref{Hee}) with Eq. (\ref{eps}) and
\begin{equation}\label{H0}
\hat{H}_0^{(\rm e)}=-\sum_{j=1}^r \cos (E_x+j\eta
-\hat{v}_j).
\end{equation}
The $\imath$th term in the expansion (\ref{Hee}), $\imath\geq 1$, is a linear combination of (repeated) commutators, each involving $\imath +1$ operators (\ref{Vt}) [for example, $[\hat{A},[\hat{A},\hat{B}]]$ in Eq. (\ref{id}) involves three operators]. Thus, for $\imath = 1$, we get
\begin{eqnarray}\label{h1_def}
\epsilon \hat{H}_{1}^{(\rm e)}
=\frac{i}{2\mu}
\sum_{j=1}^{r} \sum_{j'=j}^{r} \left[\hat{O}_j,\hat{O}_{j'}\right]. 
\end{eqnarray} 

Consider first the case of the operator (\ref{Uper}) for $\eta =0$ and $r=4$, i.e., the operator (\ref{Upa4}). In this case, from the definition of $\hat{v}_j$, the sum in Eq. (\ref{H0}) can be straightforwardly performed, giving Eq. (\ref{HeQMEta0a}). Next, using Eqs. (\ref{eps}), (\ref{Vt}), (\ref{c1}), and the fact that the commutator 
$[\hat{v}_{j},\hat{v}_{j'}]$ is nonzero only for 
$(j,j')=(1,2)$, $(1,4)$, $(2,3)$, $(3,4)$, one obtains from Eq. (\ref{h1_def}) the result in Eq. (\ref{HeQMEta0b}).

In the case of the operator (\ref{Uper}) for $\eta\neq 0$ and $r>4$, let us write $r={\rm lcm}(4,\ell )=4\ell '$, where 
the integer $\ell '>1$. Clearly, one has $4/\ell =n'/\ell '$, where $n'$ and $\ell '$ are relatively prime. Then, from Eq. (\ref{H0}),
\begin{eqnarray}\label{H0e}
& & \hat{H}_0^{(\rm e)} =\frac{1}{2} \sum_{j=1}^r e^ {i(E_x+j\eta)} e^{-i\hat{v}_j}+{\rm c.c.} \notag \\
& = & \frac{1}{2}e^{iE_x}\sum_{\bar{n}=1}^4\sum_{l=0}^{\ell '-1}e^{i[(4l+\bar{n})\eta -\hat{v}_{4l+\bar{n}}]}+{\rm c.c.}  \\ 
& = & \frac{1}{2}e^{iE_x}\sum_{\bar{n}=1}^4 e^{i(\bar{n}\eta -\hat{v}_{\bar{n}})}\frac{e^{2\pi ikn'}-1}{e^{2\pi ikn'/\ell '}-1} +{\rm c.c.}, \notag
\end{eqnarray}
where we used the periodicity of $\hat{v}_{j}$ with period $n=4$ and the fact that $4\eta=2\pi kn'/\ell '$ to perform the sum over $l$. The latter is a geometric sum, equal to the ratio in the last line of Eq. (\ref{H0e}). Clearly, this ratio is identically zero for $\ell '>1$, giving Eq. (\ref{HeQMEtaNon0a}).

Finally, to get Eq. (\ref{HeQMEtaNon0b}), let $\hat{S}_{j,j'}=\sin (x_{\mathrm c}-j\eta-\hat{v}_{j'})$. 
Then, using again the periodicity of $\hat{v}_{j}$ with period $4$, the sum over $j'$ in Eq. (\ref{h1_def}) can be written as 
\begin{eqnarray}\label{sumC}
\notag
\sum_{j'=j}^{r} \hat{O}_{j'}
&=&\frac{i\mu}{2\sin(2\eta)}\sum_{j'=j}^{r} \left(\hat{S}_{j'+2,j'}-\hat{S}_{j'-2,j'}\right)\\
\notag
&=&\frac{i\mu}{2\sin(2\eta)}\left( \sum_{j'=j}^{r} \hat{S}_{j'+2,j'}-\sum_{j'=j-4}^{r-4} \hat{S}_{j'+2,j'+4}\right)\\
\notag
&=&\frac{i\mu}{2\sin(2\eta)}\left( \sum_{\bar{n}=0}^{3} \hat{S}_{r-\bar{n}+2,r-\bar{n}}-\sum_{\bar{n}=1}^{4} \hat{S}_{j-\bar{n}+2,j-\bar{n}}\right) .\\
\end{eqnarray}
Substituting Eq. (\ref{sumC}) into Eq. (\ref{h1_def}), one finds that the first sum in the last expression of Eq. (\ref{sumC}) does not contribute since it is independent of $j$ and then one can use $\sum_{j=1}^{r}\hat{O}_{j}=0$ as in Eq. (\ref{H0e}). Concerning the second sum, only the terms of $\bar{n}=1,3$ contribute to Eq. (\ref{h1_def})  
by definition of $\hat{v}_{j}$. Then, using Eq. (\ref{c1}), we obtain, after some straightforward algebra,  
\begin{eqnarray}\label{h1_sumj}
\hat{H}_{1}^{(\rm e)}
&=&\frac{1}{2\sin(2\eta)}
\left\{ \sum_{j=1}^{r} \sin(\hat{v}_{j}-\hat{v}_{j+1}-\eta) \right.\\
\notag
&-&\left. \sum_{j=1}^{r} \cos[2E_x+(2j+1)\eta ]
\sin(\hat{v}_{j}-\hat{v}_{j+1}) \right\}. 
\end{eqnarray}
The second sum in Eq. (\ref{h1_sumj}) vanishes for $\ell ' >2$ since it contains a geometric sum similar to that in Eq. (\ref{H0e}). Thus, for $\ell ' >2$ or $r>8$, only the first sum in Eq. (\ref{h1_sumj}) contributes and this can be easily shown to give Eq. (\ref{HeQMEtaNon0b}). 

\renewcommand{\theequation}{B\arabic{equation}}
\setcounter{equation}{0}
\begin{center}
\textbf{APPENDIX B}
\end{center}

In this Appendix, we consider several properties of the classical analog of the quantum system (\ref{eKHO}) under the conditions (\ref{rc}) with $\eta =0$. Some known results will be summarized and other ones will be derived in detail.

\begin{center}
\textbf{Basic classical map}
\end{center}

We summarize here and in the next section some results in Ref. \cite{bhd}. The commutation relations (\ref{cr}) are replaced, classically, by the corresponding Poisson brackets, $[\ ,\ ]/(i\hbar )\rightarrow \{\ ,\ \}$. If $H$ is the classical analog of the quantum Hamiltonian (\ref{eKHOs}) for $\eta =0$ ($E_y=0$), one has the Hamilton equation 
$\dot{x}_{\mathrm{c}}=-\partial H/\partial y_{\mathrm{c}}=0$, meaning that $x_{\mathrm{c}}$ is a constant of the motion. As mentioned in Sec. II A, we can choose this constant to be $x_{\mathrm{c}}=0$, without loss of generality. Then, for the conjugate pair $(u,v)$, the Hamiltonian $H$ with the potential (\ref{Vk}) can be regarded as that of a harmonic oscillator periodically kicked by the potential $-\kappa \cos (E_x-v)$. The Hamilton equations $\dot{u}=\partial H/\partial v$ and $\dot{v}=-\partial H/\partial u$ can be explicitly written and, after integrating them in one time period from $t=sT-0$ to $t=(s+1)T-0$, one obtains the one-period Poincar\'{e} map for the variables $u_s=u(t=sT-0)$ and $v_s=v(t=sT-0)$:  
\begin{equation}\label{cMh}
M_{\gamma}:\ z_{s+1}=[z_s-\kappa \sin (E_x-v_s)]e^{-i\gamma},
\end{equation}
where $z_s=u_s+iv_s$ and $\gamma =\omega T=T$. Assuming the condition (\ref{rc}) for $\gamma=\pi /2$, we obtain the fourth iterate of the map (\ref{cMh}): 
\begin{equation}\label{cMhb}
M_{\gamma}^4:\ z_{s+4}=z_s-\kappa\sum_{j=0}^{3}\sin \big(E_x
-v_{s+j}\big)i^j. 
\end{equation}
The map (\ref{cMhb}) is the smallest iterate of the map (\ref{cMh}) that is a near-identity map for small $\kappa$. In addition, the map (\ref{cMhb}) is translationally invariant in the $(u,v)$ phase space with a $2\pi \times 2\pi$ unit cell (the ``fundamental domain"). One may consider (\ref{cMhb}) as the basic map of the system, analogous to the evolution operator (\ref{Upa4}) (for $x_{\mathrm{c}}=0$). 

\begin{center}
\textbf{Classical effective Hamiltonian}
\end{center}

Now we introduce a classical effective Hamiltonian $H^{(\mathrm{e})}$ as follows. From the basic map (\ref{cMhb}) with $s=0$, one has  
\begin{equation}\label{He_flow}
\frac{z_4-z_0}{\kappa} =-\sum_{j=0}^{3}\sin (E_x-v_{j})i^j. 
\end{equation}
By regarding the small value of $\kappa$ as a small time step $\Delta t$, the left hand side of Eq. (\ref{He_flow}) gives an approximation of a time derivative $\dot{z}$ of $z= u+iv$. If the right hand side of Eq. (\ref{He_flow}) is written as $\partial H^{(\mathrm{e})}/\partial v - i \partial H^{(\mathrm{e})}/\partial u$, a trajectory $z$ of the basic map is approximately given by the solution of a Hamiltonian flow. An expansion of $H^{(\mathrm{e})}$ in powers of $\kappa$, that is consistent with the quantum expansion (\ref{Hee}) in powers of $\epsilon$ [Eq. (\ref{eps})] in the classical limit of $\hbar\rightarrow 0$ ($\epsilon \rightarrow \kappa /2$), is 
\begin{equation}\label{He_expand}
H^{(\mathrm{e})}(u,v)=\sum_{\imath =0}^{\infty} 
\left(\frac{\kappa}{2}\right)^{\imath} H_{\imath}^{(\mathrm{e})}(u,v).
\end{equation}
The first terms of Eq. (\ref{He_expand}) can be explicitly derived as explained below:   
\begin{eqnarray}\label{HeEta0}
\notag
H_0^{(\mathrm{e})}(u,v)&=&-2\cos(E_x)[\cos(u) + \cos(v)],\\
\notag
H_1^{(\mathrm{e})}(u,v)&=& -\cos(u+v)+\cos(2E_x)\cos(u-v)\\
\notag
&~&-\sin(2E_x)\sin(u+v), \\ 
\notag
H_2^{(\mathrm{e})}(u,v)&=&  
\cos(E_x)[\cos(u) + \cos(v)] \\
\notag
&~&+\sin(E_x)\cos(2 E_x)[\sin(u) + \sin(v)] \\
\notag
&~&-\cos(E_x) \cos(2 E_x)[\cos(2 u)\cos(v) \\
\notag
&~&+\cos(u)\cos(2 v)] \\
\notag
&~&-\frac{1}{2}\sin(E_x)[\sin(u - 2v) - \sin(2 u - v) \\ 
&~&+\sin(2 u + v) + \sin(u + 2 v)]. 
\end{eqnarray}
The first two terms in Eq. (\ref{HeEta0}) coincide with the classical limits of Eqs. (\ref{HeQMEta0a}) and (\ref{HeQMEta0b}). Let us explain how Eqs. (\ref{HeEta0}) are derived. Equation (\ref{He_flow}) can be written as 
\begin{equation}\label{He_diff}
\frac{(u_4-u_0)+i(v_4-v_0)}{\kappa} = \delta u(u_0,v_0)+i\delta v(u_0,u_4,v_0), 
\end{equation}
where  
\begin{eqnarray}\label{dudv}
\notag
&&\delta u(u,v)=\sin(v-E_x) \\
\notag
&&~~~~+\sin\Big(v+E_x
-\kappa\sin\big[ u+\kappa\sin(v-E_x)
+E_x\big] \Big),\\
\notag
&&\delta v(u,\tilde{u},v)= -\sin(\tilde{u}-E_x)\\
&&~~~~~~~~~~~~~~~~~-\sin\big[ u+\kappa\sin(v-E_x)
+E_x\big]. 
\end{eqnarray} 
We evaluate the right-hand side of Eq. (\ref{He_diff}) at the middle point, i.e. $\bar{u}\equiv (u_4+u_0)/2$ and $\bar{v}\equiv (v_4+v_0)/2$. From Eq. (\ref{He_diff}), one can write $u_4$, $u_0$, and $v_0$ as $u_4=\bar{u}+\kappa\delta u(u_0,v_0)/2$, $u_0=\bar{u}-\kappa\delta u(u_0,v_0)/2$, and $v_0=\bar{v}-\kappa\delta v(u_0,u_4,v_0)/2$. Using these three equations, one can determine self-consistently, after lengthy but straightforward calculations, $\delta u(u_0,v_0)$ and $\delta v(u_0,u_4,v_0)$ as functions of $\bar{u}$ and $\bar{v}$, at each order of $\kappa$. After denoting $\bar{u}$ and $\bar{v}$ as $u$ and $v$, respectively, one finally finds that the right-hand side of Eq. (\ref{He_diff}) can be written as  
$\partial H^{(\mathrm{e})}/\partial v - i \partial H^{(\mathrm{e})}/\partial u$ with $H^{(\mathrm{e})}$ given by Eqs. (\ref{HeEta0}). Note that for $E_x=\pi/2$, $H_0^{(\mathrm{e})}$ in Eq. (\ref{HeEta0}) vanishes and then $H_1^{(\mathrm{e})}$ is the leading term. This fact is connected with the drastic changes of the phase-space structure as $E_x$ is varied, see details below. 
 
\newpage 
 
\begin{center}
\textbf{Topological changes of the separatrix and the phase-space structure}
\end{center}

\begin{figure}[tbp]
\includegraphics[width=6.5cm,trim = {1cm 0.2cm 0cm 0cm}]{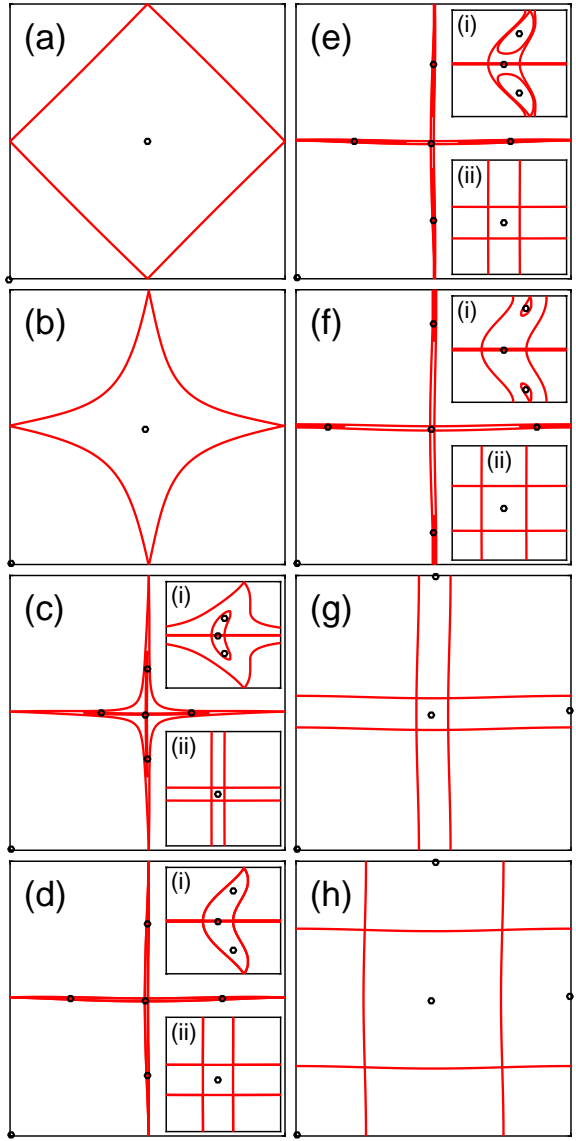}
\caption{(Color online) Phase spaces of the basic map (\ref{cMhb}) in the fundamental domain $0\le u, v <2\pi$ for $\kappa=0.05712$ and several values of $2E_x/\pi $ . Shown are separatrices (red solid lines) and stable fixed points (small black open circles). (a) $2E_x/\pi=0$; (b) $2E_x/\pi=0.98$; (c) $2E_x/\pi=0.98182$; (d) $2E_x/\pi=0.98182806174$; (e) $2E_x/\pi=0.981829$; (f) $2E_x/\pi=0.98184$; (g) $2E_x/\pi=0.983$; (h) $2E_x/\pi=1$. For (c)-(f), magnified plots are shown in the insets covering the regions (i) $3\le u \le 3.25$, $0\le v <2\pi$ and (ii) $3\le u,v \le 3.25$. The separatrix changes from the diamond type [(a) and (b)] to the cross one [(g) and (h)] by the topological changes of the phase space successively occurring when $E_x$ is varied between $E_x=E_x^*$ and $E_x^\prime$ ($2E_x^*/\pi=0.981818...$ and $2E_x^\prime/\pi=0.981848...$), as shown in plots (c)-(f). See text for details.}
\label{map-separatrix}
\end{figure}

We present here, for the convenience of the reader, a very detailed derivation of results in Ref. \cite{prk}. A separatrix of the basic map (\ref{cMhb}) is an orbit connecting unstable fixed points of the map, i.e., points $(u_0,v_0)$ satisfying $u_4=u_0$ and $v_4=v_0$ and unstable under small perturbations $(\delta u_0,\delta v_0)$. For example, the ``diamond" separatrix in Fig. 2 is the red line connecting the unstable fixed points at the ``corners". We show that the number and positions of the unstable fixed points change as $E_x$ is varied, leading to topological changes of the separatrix and the phase-space structure. Let us derive expressions for the exact positions of the fixed points. From Eq. (\ref{He_diff}), the conditions $u_4=u_0=u$ and $v_4=v_0=v$ imply that $\delta u(u,v)=0$ and $\delta v(u,u,v)=0$. Using Eqs. (\ref{dudv}), one then gets 
\begin{eqnarray}
\notag
\sin\left[u+\frac{\kappa}{2}\sin(v-E_x)\right]
\cos\left[E_x+\frac{\kappa}{2}\sin(v-E_x)\right]&=&0,\\
&&\label{fp_coupled_1}\\
\notag
\sin\left[v+\frac{\kappa}{2}\sin(u-E_x)\right]
\cos\left[E_x+\frac{\kappa}{2}\sin(u-E_x)\right]&=&0,\\
&&\label{fp_coupled_2}
\end{eqnarray} 
Equation (\ref{fp_coupled_1}) is satisfied under one of two possible 
conditions:
\begin{eqnarray}
\label{fp_condition1}
&&u+\frac{\kappa}{2}\sin(v-E_x)=m_1 \pi, \\  
\label{fp_condition2}  
&&E_x+\frac{\kappa}{2}\sin(v-E_x)=(2m_2+1)\frac{\pi}{2},  
\end{eqnarray} 
where $m_1$ and $m_2$ are integers. For Eq. (\ref{fp_coupled_2}), one has two similar conditions:
\begin{eqnarray}\label{fp_condition3}
&&v+\frac{\kappa}{2}\sin(u-E_x)=m_3 \pi, \\ 
\label{fp_condition4}
&&E_x+\frac{\kappa}{2}\sin(u-E_x)=(2m_4+1)\frac{\pi}{2},  
\end{eqnarray} 
with integers $m_3$ and $m_4$. Due to the $2\pi$-translational invariance in $u$ and $v$, one can restrict both $m_1$ and $m_3$ to $0$ or $1$ for small $\kappa$. 

For $0 \le E_{x} < E_x^*$, where $E_x^*= (\pi-\kappa)/2$,  Eqs. (\ref{fp_condition2}) and (\ref{fp_condition4}) do not have any solution for $u$ and $v$, so that the fixed points come only from Eqs. (\ref{fp_condition1}) and (\ref{fp_condition3}). 
These equations give two stable fixed points $S_1$ and $S_2$ for $(m_1,m_3)=(0,0)$ and $(m_1,m_3)=(1,1)$, respectively, 
and two unstable ones $U_1$ and $U_2$ for $(m_1,m_3)=(1,0)$ and $(m_1,m_3)=(0,1)$, respectively. The (linear) stability or instability of a fixed point is determined from the map (\ref{cMhb}) by its linear-stability (derivative) matrix $DM_{\gamma}^4$ evaluated at that point. The point is stable if Tr($DM_{\gamma}^4)<2$ and is unstable if Tr($DM_{\gamma}^4)>2$. The positions of the four points above are   
\begin{eqnarray}
\label{PS1}
P_{S_1}&=&\big( d_1(E_x;\kappa),d_1(E_x;\kappa) \big), \\
\label{PS2}
P_{S_2}&=&\big( \pi + d_1(E_x;-\kappa),
\pi + d_1(E_x;-\kappa) \big), \\  
\label{PU1}
P_{U_1}&=&\big( \pi - d_2(E_x;-\kappa),
2\pi - d_2(E_x;\kappa) \big) 
\end{eqnarray}
[$P_{U_2}$ is the mirror-symmetric point of $P_{U_1}$ relative to the line $u=v$], where $d_1$ and $d_2$ satisfy  
\begin{eqnarray}
\label{d1_def}
&&d_1(E_x;\kappa)+\frac{\kappa}{2}\sin\left[ d_1(E_x;\kappa) - 
E_x \right] = 0,\\ 
\notag
\label{d2_def}
&&d_2(E_x;\kappa)+\frac{\kappa}{2}\sin\left[ \frac{\kappa}{2}\sin\left[d_2(E_x;\kappa) + 
E_x \right]- 
E_x \right] = 0. \\  
\end{eqnarray}
The quantities $d_1$ and $d_2$ are generally small since their leading term in $\kappa$ is $(\kappa/2)\sin(E_x)$; in particular, $d_1(0;\kappa)=d_2(0;\pm \kappa)=0$. The unstable fixed points $U_1$ and $U_2$, as well as the equivalent points by translational invariance, are connected by the diamond separatrix, see Figs. 2, \ref{map-separatrix}(a), and \ref{map-separatrix}(b). 

At $E_{x} = E_x^*=(\pi-\kappa)/2$, four stable fixed points $S_3^{(\pm)}$ and $S_4^{(\pm)}$ and four unstable fixed points $U^{(\pm,\pm)}$ born at the same point $P^*=(\pi-\kappa/2,\pi-\kappa/2)$. More precisely, $S_3^{(\pm)}$, $S_4^{(\pm)}$, and $U^{(\pm,\pm)}$ emerge, respectively, as solutions of Eqs. [(\ref{fp_condition1}),(\ref{fp_condition4})] with $(m_1,m_4)=(1,0)$, [(\ref{fp_condition2}),(\ref{fp_condition3})] with $(m_2,m_3)=(0,1)$, 
and [(\ref{fp_condition2}),(\ref{fp_condition4})] with $(m_2,m_4)=(0,0)$. The positions of these points are
\begin{eqnarray}
\label{S3}
P_{S_3^{(\pm)}}&=&\left( \pi + d_{+} (\xi;\kappa), \tilde{d}_{\pm}(\xi;\kappa) \right),\\
\label{Upmpm}
P_{U^{(\pm,\pm)}}&=&\left( \pi + d_\pm (\xi;\kappa), \pi + d_{\pm}(\xi;\kappa) \right) 
\end{eqnarray}
[$P_{S_4^{(\pm)}}$ are the mirror-symmetric points of $P_{S_3^{(\pm)}}$ relative to the line $u=v$]. Here $\xi$ is introduced as   
\begin{eqnarray}
\label{xi_def}
E_x &=& \pi/2 - \xi\kappa/2,
\end{eqnarray}	
and four functions $d_{\pm}$, $\tilde{d}_{\pm}$ are defined by  
\begin{eqnarray}
\label{dpm}
d_\pm (\xi;\kappa) &=& -\xi \frac{\kappa}{2}  \pm \arccos (\xi), \\ 
\notag
\label{dtilde}
\tilde{d}_{\pm}(\xi;\kappa) &=& -\xi\frac{\kappa}{2}
\pm \arccos \left[ \frac{2}{\kappa} \arccos(\xi) - \xi \right] \mathrm{mod} (2\pi).\\     
\end{eqnarray}	
Note that $\tilde{d}_{\pm}$ in Eq. (\ref{dtilde}) is real only in the interval $1 \ge \xi \ge \xi^\prime $, corresponding to   
$E_x^* \le E_{x} \le E_x^\prime$ by Eq. (\ref{xi_def}), where $E_x^\prime$ satisfies 
\begin{eqnarray}\label{x'_simple}
E_x^\prime + \frac{\kappa}{2}\sin\left(E_x^\prime - \frac{\kappa}{2} \right) = \frac{\pi}{2}. 
\end{eqnarray}
One has the expansion $E_x^\prime=E_x^* + \kappa^3 /4 + \cdots$. Therefore, within the short interval $E_x^* \le E_{x} \le E_x^\prime$ there exist six stable (unstable) fixed points, i.e. $S_1$, $S_2$, $S_3^{(\pm)}$, and $S_4^{(\pm)}$ ($U_1$, $U_2$, and $U^{(\pm,\pm)}$) in the fundamental domain. 

As $E_x$ is slightly increased beyond $E_x^*$, the points $S_3^{(\pm)}$, $S_4^{(\pm)}$, and $U^{(\pm,\pm)}$ leave the point $P^*$ above and form a ``cross" separatrix inside the diamond one, see Fig. \ref{map-separatrix}(c) and insets. As $S_3^{(+)}$ ($S_3^{(-)}$) leaves $P^*$, it moves down (up) quickly along the line $u=\pi$. Also, $S_4^{(+)}$($S_4^{(-)}$) moves right (left) quickly along the line $v=\pi$. Then, the cross separatrix expands. At some value of $E_x$, the cross separatrix merges with the diamond one (formed by $S_1$, $S_2$, $U_1$, and $U_2$), see Fig. 
\ref{map-separatrix}(d) and insets. Immediately after this merging, the diamond separatrix disappears. After that, the cross separatrix extends over the whole fundamental domain and islands appear inside the separatrix, with ``corners" at the points $U_1$ and $U_2$; see Fig. \ref{map-separatrix}(e) and insets. As $E_x$ approaches $E_x^\prime$, both $S_3^{(\pm)}$ ($S_4^{(\pm)}$) get close to $U_1$ ($U_2$) and the islands above shrink, see Fig. \ref{map-separatrix}(f) and insets.    
At $E_x=E_x^\prime$, $S_3^{(\pm)}$ ($S_4^{(\pm)}$) finally merge with $U_1$ ($U_2$) at $P^\prime = \left( \pi + \kappa/2, 3\pi/2 + E_x^\prime \right)$ (This can be checked by using, for $S_3^{(\pm)}$, Eq. (\ref{S3}) with $d_{+}(\xi^\prime;\kappa)=\kappa/2$ and $\tilde{d}_{\pm}(\xi^\prime;\kappa)=2\pi-\xi^\prime\kappa/2=3\pi/2+E_x^\prime$, and, for $U_1$, 
Eq. (\ref{PU1}) with $E_x^\prime + d_2\left(E_x^\prime; \kappa \right) = \pi/2$ from Eqs. (\ref{d2_def}) and (\ref{x'_simple})). Then, $S_3^{(\pm)}$ ($S_4^{(\pm)}$) disappear and $U_1$ ($U_2$) changes from unstable to stable fixed point, see Fig. \ref{map-separatrix}(g). We denote this new stable point by $\bar{S}_1$ ($\bar{S}_2$). The positions of $\bar{S}_1$ and $\bar{S}_2$ are still given by $P_{U_1}$ in Eq. (\ref{PU1}) and its mirror-symmetric point $P_{U_2}$, respectively. 

Thus, for $E_x^\prime < E_x \le \pi/2$, there are four stable fixed points $S_1$, $S_2$, $\bar{S}_1$, and $\bar{S}_2$, 
and four unstable fixed points $U^{(\pm,\pm)}$, which characterize the cross separatrix in the fundamental domain; see Figs. \ref{map-separatrix}(g) and \ref{map-separatrix}(h). For $E_x=\pi/2$, using $d_\pm (0;\kappa)=\pm\pi/2$ from Eq. (\ref{dpm}), one finds that the points $U^{(\pm,\pm)}$ in Eq. (\ref{Upmpm}) are positioned simply at $\left(\pi \pm \pi/2, \pi \pm \pi/2 \right)$, see Fig. \ref{map-separatrix}(h).  

\begin{center}
\textbf{Energy of separatrix}
\end{center}

The exact orbits of the map (\ref{cMhb}) are orbits of a time dependent (periodic) Hamiltonian, so that their energy is not exactly conserved. However, for small $\kappa$, the map (\ref{cMhb}) is approximately described by an effective 
time-independent Hamiltonian (\ref{He_expand}) and one can then associate an effective energy with an orbit of this Hamiltonian. This energy is just the value of the Hamiltonian at any point of the orbit. In particular, the energy of the diamond separatrix for $0 \le E_{x} < E_x^*$ (see above) can be calculated as $H_{\rm sep}^{(\mathrm{e})}=H^{(\mathrm{e})}(U_{1,2})$, where $U_1$ and $U_2$ are the unstable fixed points connected by this separatrix. The exact position of $U_1$ is given by Eq. (\ref{PU1}) and the function $d_2(E_x;\kappa)$ in Eq. (\ref{d2_def}) can be expanded in $\kappa$ as $(\kappa/2)\sin(E_x)+{\mathcal O}(\kappa^2)$. After substituting the corresponding expansion for Eq. (\ref{PU1}) into Eq. (\ref{He_expand}) with Eq. (\ref{HeEta0}), one finds that $H_{\rm sep}^{(\mathrm{e})}$ starts from a linear term in $\kappa$, coming only from $(\kappa/2)H_1^{(\mathrm{e})}$. There is not a $\kappa^2$ term since $H_0^{(\mathrm{e})}$, $(\kappa/2)H_1^{(\mathrm{e})}$, and $(\kappa/2)^2H_2^{(\mathrm{e})}$ do not include it. One then gets formula (\ref{Es}) for $H_{\rm sep}^{(\mathrm{e})}$. The same result is obtained, of course, by using $P_{U_2}$ instead of $P_{U_1}$.


\begin{thebibliography}{99}
\bibitem{note} Our notation here is different from that used in Refs. \cite{d2,dk,dd1}.
\bibitem{note1} In fact, assuming that there is a second solution $(\sigma ',\mu ')$ of Eq. (\ref{DiophSumIrr}) at fixed $\zeta$, one gets $\hbar_{\mathrm{s}}= (\sigma '-\sigma)/(\mu -\mu ')$, contradicting the fact that $\hbar_{\mathrm{s}}$ is irrational.
\end{thebibliography}
\end{document}